\documentclass[prd,aps,nofootinbib,floatfix,11pt]{revtex4}
\usepackage{amsmath,graphicx,epsfig,amssymb,dsfont,mathtools,}
\usepackage[usenames]{color}
\usepackage{ulem} 
\usepackage{bigstrut}
\usepackage{slashed}
\usepackage{multirow}
\usepackage{subfigure}
\usepackage{amsmath}

\allowdisplaybreaks

\begin{document}
\title{Analysis of molecular state ${{\eta}_cD^*}$ and ${J/\psi D^*}$ in the effective Lagrangian approach}
\author{Na Li}
\email{d202580107@hust.edu.cn}
\affiliation{School of Material Science and Physics,China University of Mining and Technology, Xuzhou  221000, China }
\affiliation{School of physics, Huazhong University of Science and Technology, Wuhan 430074, China }

\author{Ye Xing}
\email{xingye_guang@cumt.edu.cn}
\affiliation{School of Material Science and Physics,China University of Mining and Technology, Xuzhou  221000, China }
\author{Jing-Rui Shi}
\affiliation{School of Material Science and Physics,China University of Mining and Technology, Xuzhou  221000, China }

\begin{abstract}
 In this work, we  investigate the production and decay of the molecular states $cc\bar c\bar q$ with $J^P=1^+$ using the phenomenological analysis and effective Lagrangian approach. Based on an SU(3) flavor symmetry analysis to identify golden channels, we further explore the dynamics of these processes under the molecular assumptions of ${\eta_c D^*}$ and ${J/\psi D^*}$. Our results indicate that the production branching ratio from $B_c$ meson is sizable, it can reach the order of $10^{-4}$ for the molecular configuration ${{\eta}_cD^*}$, and $10^{-5}$ for molecule ${J/\psi D^*}$. In addition, we find that the decay widths of the two molecular configurations ${{\eta}_cD^*}$ and  ${J/\psi D^*}$ are not significant, being at level of $\cal{O}$($\rm {MeV}$).
\end{abstract}

\maketitle

\section{Introduction}
Since the LHCb Collaboration found the strange-charm four-quark state candidate $X_{0,(1)}(2900)$ and $T_{c\bar{s}}^{0(++)}(2900)$~\cite{LHCb:2020pxc,LHCb:2020bls,LHCb:2022sfr,LHCb:2022lzp}, the open-charmed four-quark states have attracted increasing attention~\cite{Qin:2022nof,Dong:2020hxe,Chen:2016qju,Guo:2017jvc,Liu:2019zoy,Cao:2023rhu,Mai:2022eur,Meng:2022ozq,Ortega:2020tng,Huang:2023jec,Lebed:2023vnd,Zou:2021sha,Du:2021fmf,Liu:2024uxn,Johnson:2024omq}. Subsequently, the doubly charmed four-quark state candidate $T_{cc}^{+}(3875)$, with the mass near $(D^{+}D^0)$ threshold,~\cite{LHCb:2021vvq,LHCb:2021auc} was reported in 2021. In the case of the fully charmed four-quark states, LHCb observed the broad candidate structure that ranged from $6.2$ to $6.8$ GeV and a narrow one located around $6.9$ GeV in the $J/\Psi J/\Psi$ channel~\cite{LHCb:2020bwg}. These discoveries suggest that the heaviest open-charm four-quark state with triply charmed flavor should be possible. Although there is not yet sufficient experimental evidence for the existence, it has been many preliminary theoretical research to explore the properties of the exotic state at present~\cite{Chen:2020eyu,Zhu:2023lbx,Weng:2021ngd,Jiang:2017tdc,Mutuk:2023yev,Yang:2024nyc,Lu:2021kut,Xing:2019wil,Liu:2019mxw}. In particular, regarding the mass spectrum of triply charmed four-quark states,  Ref.~\cite{Yang:2024nyc} employed a quark-based model in conjunction with the Gaussian Expansion Method (GEM) and the Complex Scaling Method (CSM) to identify a triply heavy tetraquark state within the energy range of 5.6-5.9 GeV. In Ref.\cite{Jiang:2017tdc}, the authors estimate the mass of the $cc\bar{c}\bar{q}$ tetraquark state with quantum numbers $J^{P}=1^{+}$ to be $5.1\pm0.2$ GeV using the operator product expansion (OPE). Furthermore, regarding the internal structure of states containing three heavy quarks, analogous to fully charmed four-quark states, conventional analyses disfavor a molecular configuration because the exchange of a single long-range light meson between two charmonium is suppressed. However, using heavy meson exchange forces~\cite{Liu:2023gla,Liu:2024pio} or two light mesons exchange force~\cite{Dong:2021lkh}, the authors considered the possibility of fully heavy four-quark molecular states. If such interactions do play an important role, the triply charmed four-quark molecular states should also be possible.
Nevertheless, whether a stable or resonant triple charm four-quark state exists, and if so, whether its properties manifest as those of a compact tetraquark or an extended hadronic molecule, remains an open and fundamental question in hadron physics. Regardless of the ongoing theoretical debate, there is strong motivation for a systematic investigation of its potential phenomenological signatures, which can yield concrete theoretical predictions to provide direct guidance for experimental searches.

We focus on the production and decay properties of the molecular states $cc\bar c\bar q$ in the phenomenological analysis and effective Lagrangian approach. The SU(3) flavor symmetry analysis~\cite{Meng:2020ihj,Yang:2020nrt,Yan:2021tcp,Baru:2021ddn,Zhai:2022ied,Li:2023kcl,Xing:2019hjg,Hu:2017dzi} and the effective Lagrangian approach~\cite{Isgur:1989js,Lipkin:1996ny,Li:1996yn,Han:2021azw,Chen:2015iqa}, which have been successfully applied to heavy meson and baryon systems, can accordingly be used to investigate the  molecular states $cc\bar c\bar q$. With the assumption of molecular state, then the discussion of dynamics is possible. The hypothesis of hadronic molecule is popular, such as the four-quark candidate $\psi$(4040) is considered as $D^{*}$$D^{*}$ hadronic molecule~\cite{DeRujula:1976zlg,Voloshin:1976ap}. The $Y(4220)$, $Z_c(3900)$, and $X(3872)$ states are interpreted as ${\bar D}_1 D$, ${\bar D}^* D$, and ${\bar D}^* D$ molecular states respectively~\cite{Liu:2024ziu,Peng:2023lfw}. Moreover, the open-charm four-quark candidate $T_{cc}^{+}(3875)$ is discussed under the molecular state $(D^{+}D^{*0})$~\cite{LHCb:2021auc,Meng:2022ozq}. Accordingly, in this work we adopt the generic schemes for the triply charmed molecule with $J^P=1^+$, pseudoscalar-vector $\eta_c D^{*}$($PV$) and vector-vector ${J/\psi D^{*}}$($VV$)  configurations. The work proceeds following the standard procedure of the effective Lagrangian approach. We firstly establish a gauge invariant phenomenological Lagrangian to describe the interactions between molecular states and their components, whose associated coupling constants can be determined by the bare compositeness condition. A direct calculation depending on the Lagrangian at the hadronic level can then be carried out. To facilitate the calculation, we primarily consider the leading order triangle diagram contribution to the strong decay process involving $c\bar c$ annihilation in $J/\Psi D^*$ or $\eta_c D^*$ systems, as this process is suppressed by the OZI rule and the heavy charm quark mass.

This paper is organized as follows.  In Sec.\,\ref{The triply charmed four-quark state}, we concisely analyze the properties of the molecular states ${{\eta}_cD^*}$ and  ${J/\psi D^*}$, including their production and decay processes. The numerical results and corresponding discussions are presented in Sec.\,\ref{Numerical analysis and discuss}. A summary of the work is provided in the final section.
\section{molecular state: ${J/\psi D^{*}}$ and ${\eta_c D^{*}}$}
\label{The triply charmed four-quark state}
We focus on the possible S-wave molecular configurations with $J^P = 1^+$ for the four-quark state $T_{cc\bar{c}\bar{q}}$.
The possible molecular states can be vector-vector configurations $T_{J/\psi D^{*}}$($VV$) and vector-pseudoscalar configurations $T_{\eta_c D^{*}}$($PV$). The effective Lagrangian describing the couplings between a molecular state and its constituents are given by~\cite{Dong:2008gb},
\begin{eqnarray}
	&&{\cal L}_{T_{\text{\tiny $\eta_c D ^*$}}} (x) = g_{\text{\tiny $T_{\text{\tiny $\eta_c D^*$}}$}} \, T_{\eta_c D^*}^{\mu}(x)
	\int dy \, \Phi(y^2) \, \eta_c(x+{\omega}_1 y) \, D^*_{\mu}(x-{\omega}_2 y),\,\nonumber\\
	&&{\cal L}_{T_{\text{\tiny ${J/\psi D^*}$}}}(x)  =ig_{\text{\tiny $T_{\text{\tiny ${J/\psi D^*}$}}$}} \, \varepsilon_{\mu \nu \alpha \beta}    \partial^{\mu}T_{J/\psi D^*}^{\nu}(x)
	\int dy \, \Phi(y^2) \, J/\psi^{\alpha}(x+{\omega}_1 y) \, {D^*}^{\beta}(x-{\omega}_2 y).\,
\end{eqnarray}
Among them, the Fourier transform of the correlation function is defined as~\cite{Branz:2007xp,Li:2023kcl},
 \begin{eqnarray}
	\Phi(y^2) \, = \, \int\!\frac{d^4p}{(2\pi)^4}  \,
	e^{-ip y} \, {\widetilde{\Phi}}(-p^2) \quad \text{with}\quad \widetilde\Phi(p_E^2)
	\doteq \exp( - p_E^2/\Lambda_{T}^2).
\end{eqnarray}
The coupling constants can be further obtained by the renormalization of mass operator, which can be described by the self-energy diagram in Fig.\,\ref{fig:1}.A,
\begin{eqnarray}
\label{mass operator}
	&&\Sigma_{T}(p^2)=\frac{g^2_{T_{\text{\tiny${{\eta_c D^*}}$}}}}{3} \int \frac{{\rm d^{4}}k}{(2{\pi})^4i}\,\widetilde\Phi^2(-(k-pw)^2)\frac{-3+\frac{k^2}{m^2_{D^*}}-\frac{(k\cdot p)^2}{m^2_{T_{\eta_c D^*}}}}{(k^2-m^2_{D^*})((p-k)^2-m^2_{\eta_c})},\nonumber\\
&&\Sigma^{\prime}_{T}(p^2)=\frac{g^2_{T_{\text{\tiny${J/\psi D^*}$}}}}{3} \int \frac{{\rm d^{4}}k}{(2{\pi})^4i}\,\widetilde\Phi^2(-(k-pw)^2)\frac{{\mathcal{A}}}{(k^2-m^2_{J/\psi })((p-k)^2-m^2_{ D^*})},
\end{eqnarray}
where
\begin{eqnarray}
{\mathcal{A}}\!\!&=\!\!&\frac{(-2k \!\cdot\! p+k^2+p^2)(k \!\cdot\! p)^2}{{m^2_{J/\psi}}{m^2_{D^*}}}-\frac{2(k\!\cdot\! p)^2}{{m^2_{J/\psi}}}
-\frac{2(k \!\cdot\! p-k^2)(p^2-k \!\cdot\! p)(k \!\cdot\! p)}{{m^2_{J/\psi}}{m^2_{D^*}}}+\frac{k^2(p^2-k \!\cdot\! p)^2}{{m^2_{J/\psi}}{m^2_{D^*}}}-6 p^2\nonumber\\
&\!\!+\!\!&\frac{p^2(k \!\cdot\! p-k^2)^2}{{m^2_{J/\psi}}{m^2_{D^*}}}
-\frac{k^2 p^2(k^2-2k \!\cdot\! p+p^2)}{{m^2_{J/\psi}}{m^2_{D^*}}}+\frac{2 k^2 p^2}{{{m^2_{J/\psi}}}}\!-\!\frac{2(p^2-k \!\cdot\! p)^2}{{m^2_{D^*}}}+\frac{2 p^2(k^2-2k \!\cdot\! p+p^2)}{{m^2_{D^*}}}.
\end{eqnarray}
According to the Weinberg compositeness condition~\cite{Weinberg:1962hj},
\begin{eqnarray}\label{eq3}
	Z_{T} = 1 -  \frac{d\Sigma^{(\prime)}_{T}}{dp^2}\bigg|_{p^2= m^2_{T}} =0 \,.
\end{eqnarray}
We derive the expressions for the coupling constants $g_{T_{\text{\tiny $\eta_c D^*$}}}$ and $g_{T_{\text{\tiny $J/\psi D^*$}}}$, and adopt $\Lambda_T$ a range of 0.8 to 1.2 GeV in the numerical analysis. These expressions are provided in Appendix \,\ref{sec:production}.
\subsection{ phenomenological analysis}\label{sec:phenomenological}
In this section, we first employ a phenomenological analysis to investigate the production and decay of the molecular states. A convenient phenomenological tool is the SU(3) flavor symmetric analysis. In flavor space, light meson can be grouped into one SU(3) octet, and the triple-charm molecular state forms an SU(3) triplet, which can be represented as $T_{cc\bar c \bar q} = \left( T^{0}_{cc \bar c \bar u} ,  T^{+}_{cc \bar c \bar d}  , T^{+}_{cc \bar c \bar s} \right)$. Similarly charmed meson form an triplet $D = ( D^0 , D^+ , D_s^+)$. The production from $B_c$ meson must proceed via the weak transitions $\bar b\to  c\bar c \bar s/\bar d$. Within the framework of SU(3) flavor symmetry analysis, the effective Hamiltonian can be written as,
\begin{eqnarray}
{\mathcal{H}}= a_{1}{B}_{c}(H_{3})^{i}({T_{cc\bar c\bar q}})_{j}{V}^{j}_{i}, \ \text{with}\  (H_{3})^{2}=V_{cd}^{*}, (H_{3})^{3}=V_{cs}^{*}.
\end{eqnarray}
Here, $V$ denotes the vector meson field, the coefficient $a_{1}$ parameterizes the nonperturbative effects, and $(H_3)^i$ represents the weak transition vertex. By expanding this Hamiltonian, we obtain all possible production processes, which are listed in Tab.\,\ref{tab:production} in Appendix\,\ref{sec:production}. Taking into account the CKM matrix elements and the experimental detection efficiency,  we select three golden channels,
\begin{eqnarray}
B_c^+\to   {{\overline K}^{*0}}   T_{cc\bar c\bar s}^{+} , \,\,\,
B_c^+\to   {K^{*+}}   T_{cc\bar c \bar u}^{0} , \,\,\,
B_c^+\to  {K^{*0}}   T_{cc\bar c\bar d}^{+} .
\end{eqnarray}
Neglecting phase space effects leads to simplified relations among the different channels,
\begin{eqnarray}
\label{eqn8}
&&\frac{\Gamma(B_c^+\to K^{*+}  T_{cc\bar c\bar u}^{0})}{\Gamma(B_c^+\to \overline{K}^{*0} T_{cc\bar c\bar s}^{+})}=\frac{\Gamma(B_c^+\to K^{*0}  T_{cc\bar c\bar d}^{+})}{\Gamma(B_c^+\to \overline{K}^{*0} T_{cc\bar c\bar s}^{+})}=\frac{|V_{cs}|^2}{|V_{cd}|^2}.\,\,\
\end{eqnarray}
It should be noted that $T_{cc\bar c\bar q}$ can be molecule $T_{J/\psi D^*}$ and $T_{\eta_c D^*}$ for the production processes. However, for the strong decays of the four-quark molecular state, the phenomenology of $T_{cc\bar{c}\bar{q}}$ may differ significantly between the two molecular configurations due to phase space effects. Therefore, we write the Hamiltonians of two-body and three-body decays respectively,
\begin{eqnarray}
{\mathcal{H}}\!\!&=&\!\!b_{1}({T}_{\text{\tiny{$J/\psi D^*$}}})_i D^{i}J/\psi+b_{2}({T}_{\text{\tiny{$J/\psi D^*$}}})_i{D^*}^{i}\eta_c
+c_{1}({T}_{\text{\tiny{$J/\psi D^*$}}})_i \eta_c {D}^{j}{P}^i_j+c_{2}({T}_{\text{\tiny{$J/\psi D^*$}}})_i J/\psi {D}^{j}{M}^i_j\nonumber\\
&+&\!\!b^{\prime}_1({T}_{\text{\tiny{$\eta_c D^*$}}})_i D^{i}J/\psi
+c^{\prime}_{1}({T}_{\eta_c D^*})_i \eta_c {D}^{j}{P}^i_j,
\end{eqnarray}
where $P$ denotes a pseudoscalar meson. Accordingly, we derive all possible decay processes, which are summarized in Tab.\,\ref{tab:decay}, as well as the relations among different decay channels, which are listed in Tab.\,\ref{tab:connections}.
\begin{table}
\centering
	\caption{{The decay relations between different channels for the possible four-quark molecular states $T_{J/\psi D^*}$ and $T_{\eta_c D^*}$.}}\label{tab:connections}\begin{tabular}{ll|ll}\hline\hline
$\Gamma(T^0_{\eta_c  D^{*}}\to J\!/\!\psi  D^0)$&$= \Gamma(T^+_{\eta_c  D^{*}}\to J\!/\!\psi  D^{+})$
 &$\Gamma(T^+_{\eta_c D^*}\to    D^0 \eta_c  \pi^+)$&$=2\Gamma(T^+_{\eta_c D^*}\to    D^+  \eta_c  \pi^0 )$\\
& $=\Gamma(T^{+}_{\eta_c D_s^*}\to \eta_c  D^{*+}_s)$&&$=2\Gamma(T^0_{\eta_c D^*}\to    D^0  \eta_c \pi^0 )$\\
$\Gamma(T_{J\!/\!\psi D^*}^{0}\to J\!/\!\psi  D^0)$&$=\Gamma(T_{J\!/\!\psi D^*}^{+}\to J\!/\!\psi  D^+)$
 &$\Gamma(T^+_{J\!/\!\psi D^*}\to    D^0  J\!/\!\psi  \pi^+ )$&$=2\Gamma(T^0_{J\!/\!\psi D^*}\to    D^0  J\!/\!\psi  \pi^0 )$\\
&$=\Gamma(T_{J\!/\!\psi D_s^*}^{+}\to J\!/\!\psi  D^+_s)$&&=2$\Gamma(T^+_{J\!/\!\psi D^*}\to    D^+  J\!/\!\psi \pi^0 )$\\
$\Gamma(T_{J\!/\!\psi D^*}^{0}\to \eta_c  D^{*0})$&$=\Gamma(T_{J\!/\!\psi D^*}^{+}\to \eta_c  D^{*+})$&$\Gamma(T^0_{J\!/\!\psi D^*}\to    D^+  \eta_c  \pi^- )$&$=\Gamma(T^+_{J/\psi D^*}\to    D^0  \eta_c  \pi^+ )$\\
&$=\Gamma(T_{J\!/\!\psi  D_s^*}^{+}\to \eta_c  D^{*+}_s)$&&$=2\Gamma(T^0_{J\!/\!\psi D^*}\to    D^0  \eta_c  \pi^0 )$\\
&&&$=2\Gamma(T^+_{J\!/\!\psi D^*}\to    D^+  \eta_c  \pi^0 )$
\\\hline
		\hline
	\end{tabular}
\end{table}
\subsection{The effective Lagrangian approach}
\renewcommand\thesubsection{(\roman{subsubsection})}
\subsubsection{Production of the  molecular states from $B_c$ meson }
\begin{figure}
	\includegraphics[width=0.9\columnwidth]{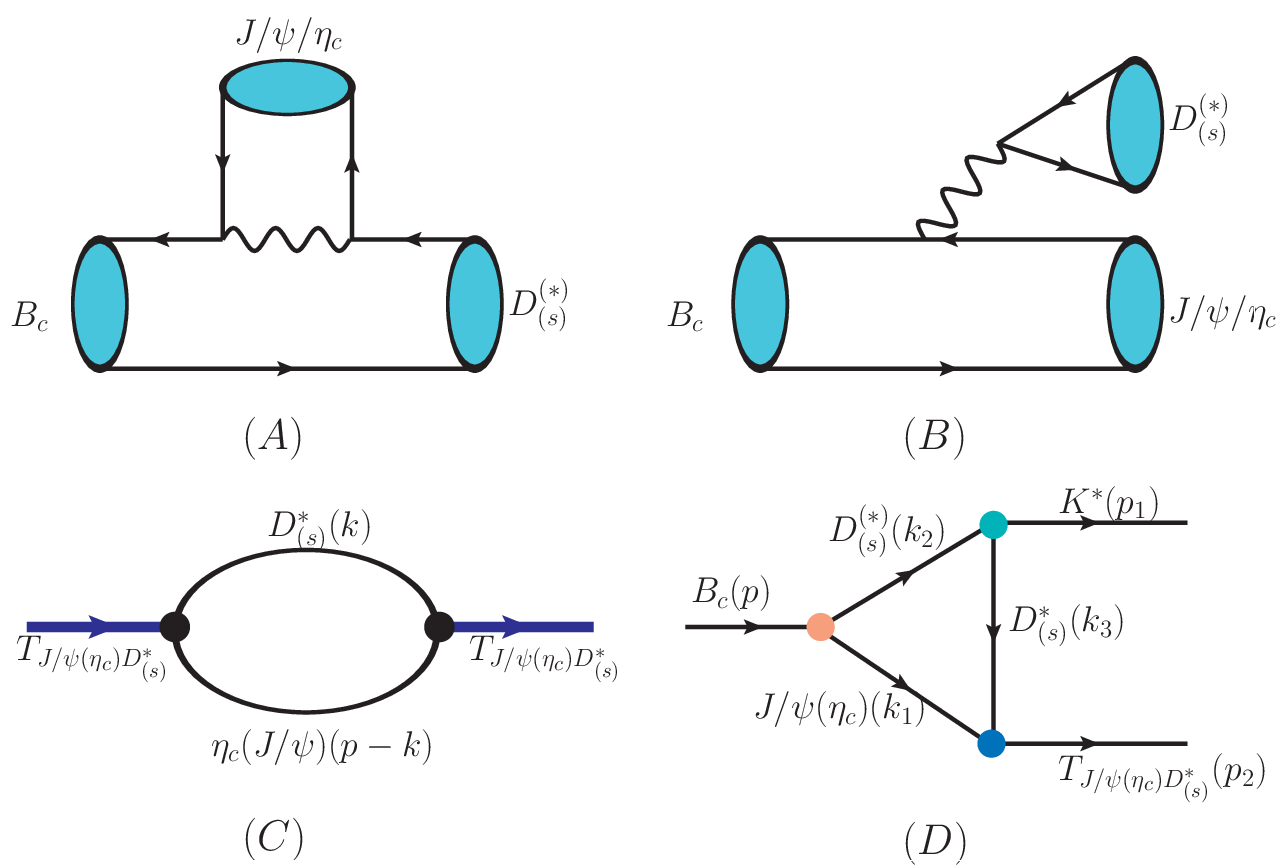}
	\caption{The internal W-emission(A) and external W-emission diagram(B) represent the weak processes of the molecular state $T_{\eta_c D^*}$ and $T_{J/\psi D^*}$ production from $B_c$ meson. Diagrams(C,D) are the self-energy and production triangle process of the molecular states $T_{J/\psi D^*}$ and $T_{\eta_c D^*}$ respectively. }\label{fig1}
\label{fig:1}
\end{figure}

Building on the SU(3) flavor symmetry analysis, we proceed to investigate the dynamics of the molecular states $T_{J/\psi D^*}$ and $T_{\eta_c D^*}$ within the effective Lagrangian approach. The production processes for these molecular states $T_{J/\psi D^*}$ and $T_{\eta_c D^*}$ can be depicted by the triangle diagram shown in Fig.\,\ref{fig:1}.D, the corresponding amplitude can be expressed as,
\begin{eqnarray}
\langle K^*T_{J/\psi(/\eta_c) D^*}|\mathcal{H}_{eff}| B_c\rangle=\sum_{\lambda}\langle K^*T_{J/\psi(/\eta_c) D^*}|\mathcal{H}_{\lambda}|M_1M_2\rangle\langle M_1M_2|\mathcal{H}_{eff}| B_c\rangle,
\end{eqnarray}
here, $\langle M_1M_2|\mathcal{H}_{eff}| B_c\rangle$ is the short distant weak decay matrix element, $M_1 M_2$ are the possible final states of $B_c$ weak decay and $H_{eff}$ is the possible weak decay Hamiltonian. In the process, W-emission diagrams shown in Fig.\,\ref{fig:1}.(A,B) are dominant for the weak decay. The matrix element $\langle K^*T_{J/\psi(/\eta_c) D^*}|\mathcal{H}_{\lambda}|M_1M_2\rangle$ describes the long distant strong coupling process, and $\mathcal{H}_{\lambda}$ denotes the corresponding effective strong interaction Hamiltonian. We show the effective weak decay Hamiltonian as,
\begin{eqnarray}
	\mathcal{H}_{eff} &=& \frac{G_F}{\sqrt{2}} V_{cb}^*V_{cq} (C_1(\mu) \mathcal(\bar b_{\alpha} c_{\beta})_{V-A}(\bar c_{\beta}q_{\alpha})_{V-A}+C_2(\mu) (\bar b_{\alpha} c_{\alpha})_{V-A}(\bar c_{\beta}q_{\beta})_{V-A})+h.c. \,,
\end{eqnarray}
where $G_F$ and $C_{1,2}(\mu)$ are the Fermi constant and Wilson coefficient respectively. The weak decay amplitude can be factorized into the product of the Wilson coefficient, nonperturbative form factor, and decay constant,
\begin{eqnarray}
	&&\mathcal{M}( B_c \to J/\psi D(D^*))=\frac{G_F}{\sqrt{2}} V_{cb}^* V_{cs} a_{1,2}  \langle J/\psi|(\bar b c)_{V-A} | B_c\rangle \langle D(D^*)|(\bar cq)_{V-A} |0\rangle \,,\\
	&&\mathcal{M}( B_c \to \eta_c D(D^*))=\frac{G_F}{\sqrt{2}} V_{cb}^* V_{cs} a_{1,2} \langle \eta_c|(\bar b c)_{V-A} | B_c\rangle \langle D(D^*)|(\bar cq)_{V-A} |0\rangle \,,
\end{eqnarray}
Here, $a_{1}$ and $a_{2}$ are the effective Wilson coefficients, with $a_{1} = C_{1} + C_{2}/N_{c}$, $a_{2} = C_{2} + C_{1}/N_{c}$. The parameter $N_{c}$ represents the number of colors. The form factors are defined as follows\cite{Dhir:2008hh},
 \begin{eqnarray}
 	\langle J\!/\!\psi(p_{\!\text{\tiny $J\!/\!\psi$}})|(\bar b c)_{V-A} | B_c(p_{\text{\tiny $B_c$}})\rangle&=&\frac{2}{m_{B_c}+m_V}\varepsilon_{\mu\nu\rho\sigma}\varepsilon^{*\nu}p_{\text{\tiny $B_c$}}^{\rho}
 	p_{\!\text{\tiny $J\!/\!\psi$}}^{\sigma} A_0(q^2)+i\varepsilon_{\mu}^*(m_{B_c}+m_V)A_1(q^2)\nonumber\\
 	&-&i\frac{\varepsilon^*\cdot q}{m_{B_c}+m_V}(p_{\text{\tiny $B_c$}}+p_{\text{\tiny $J/\psi$}})_{\mu} A_2(q^2)-i \frac{\varepsilon^*\cdot  q}{ q^2} 2 m_V  {q}_{\mu} A_3(q^2)\label{eq:formfactor1}\\\nonumber
 	&+&i \frac{\varepsilon^*\cdot {q}}{{q^2}} 2m_V {k_2}_{\mu} A_4(q^2) \,,\\
 	\langle \eta_c(p_{\text{\tiny $\eta_c$}})|(\bar b c)_{V-A} | B_c(p_{\text{\tiny $B_c$}})\rangle&=&(p_{\text{\tiny $B_c$}}+p_{\text{\tiny $\eta_c$}}-\frac{m_{B_c}^2-m_P^2}{q^2}q)_{\mu} F_1(k^2_2)+\frac{m_{B_c}^2-m_{\eta_c}^2}{q^2}{q}_{\mu} F_0(q^2).\label{eq:formfactor2}
 \end{eqnarray}
Here, $\varepsilon_{\mu}$ denotes the polarization vector of the $J/\psi$, transition momentum $q_{\mu}\equiv(p_{\text{\tiny $B_c$}}-p_{\text{\tiny $J/\psi$}})_{\mu}$.
The matrix elements between meson $D^{(*)}$ and vacuum are defined as~\cite{Cheng:1996cs}
\begin{eqnarray}
	\langle D|(\bar c q)_{V-A} |0\rangle=i f_{D} \, q_{\mu} \,, \qquad \langle D^*| (\bar cq)_{V-A}|0 \rangle=m f_{D^*} \varepsilon^*_{\mu} \,,
\end{eqnarray}
where $f_{D^{(*)}}$ is the decay constant of $D^{(*)}$ meson, $\varepsilon_{\mu}$ represents the polarization vector of $D^*$ meson.

To deal with the long distant strong coupling matrix, we introduce the following strong interaction effective Lagrangian~\cite{Shen:2019evi},
\begin{eqnarray}
	&&\mathcal{L}_{V P_1 P_2} = i g_{V P_1 P_2}^{} \left( V_{\mu} \partial^{\mu} P_1 P_2 - V_{\mu} \partial^{\mu} P_2 P_1\right),\nonumber\\
		&&\mathcal{L}_{V_1 V_2 P} = -g_{V_1 V_2 P}^{} \ \varepsilon^{\mu \nu \alpha \beta} \left(\partial_{\mu} V_{1 \nu}  \partial_{\alpha} V_{2 \beta}\right) P,\nonumber\\
		&&\mathcal{L}_{D^* D^* J/\psi} = -i g_{D^* D^* J/\psi}^{} \Bigl\{ {D^*}^{\mu}\left(\partial_{\mu} {D^*}^{\nu} {J/\psi} _{\nu} - {D^*}^{\nu} \partial_{\mu} {J/\psi} _{\nu}\right) + \left(\partial_{\mu}{D_{ \nu}^*} {D^*}^{\nu} - {D_{ \nu}^*} \partial_{\mu}{D^*}^{\nu} \right){J/\psi}^{\mu} \nonumber \\
&&\,\,\,\,\,\,\,\,\,\,\,\,\,\,\,\,\,\,\,\,\,\,\,\,\,\,\,+ {D^*}^{\mu}\left({D^*}^{\nu} \partial_{\mu} {J/\psi}_{ \nu} - \partial_{\mu} {D_{ \nu}^*}{J/\psi}^{\nu}\right) \Bigr\},
\end{eqnarray}
where $V_i$ and $P_i$ denote vector and pseudoscalar meson respectively.

Using the effective Lagrangian and form factors, the amplitudes of $B_{c}\to K^*T_{J/\psi(/\eta_c) D^*}$ in Fig.\,\ref{fig:1}(D) are given as follows,
\begin{eqnarray}
{\mathcal M}_{{T_{J/\psi D^*}}}\!\!&=&\!\!g_{\text{\tiny $T_{J/\psi D^*}$}}\int\frac{d^4k_3}{(2\pi)^4}\mathcal{F}_2(k_3^2)\Big(\frac{2}{m_{B_c}+m_{J/\psi}} \varepsilon_{\varphi \nu \rho \sigma} p^{\rho} k_1^\sigma A_0(k_2^2)+i g_{\nu \varphi} A_1(k_2^2)\nonumber\\
&&(m_{B_c}+m_{J/\psi})-\frac{i {k_2}_{\nu}}{m_{B_c}+m_{J/\psi}}(p+k_1)_{\varphi} A_2(k_2^2)-2 i m_{J / \psi} \frac{{k_2}_{\nu} {k_2}_{\varphi}}{k_2^2} A_3(k_2^2)\nonumber\\
&&+2 i \frac{{k_2}_{\nu} k_{2\varphi}}{k_2^2} m_{J / \psi} A_4(k_2^2)\Big)(-g^{\nu\beta_1}+\frac{k_1^{\nu}k_1^{\beta_1}}{m_{J/\psi}^2})(-g^{\alpha_1\beta}+\frac{k_3^{\alpha_1}k_3^{\beta}}{m_{D^*}})p_2^\mu \varepsilon^{*\nu_1}(p_2)\varepsilon^{* n}(p_1)\nonumber\\
&&\times \Big\{\frac{if_{D_s}g_{\text{\tiny $D_sD^*K^*$}} k_3^{\alpha}k_2^{\varphi} \varepsilon_{mn\alpha \beta} p_1^m }{(k_1^2-m_{J/\psi}^2)(k_3^2-m_{D^*}^2)(k_2^2-m_{D_s}^2)}+\frac{m_{D_s^*}f_{D_s^*}g_{\text{\tiny$D_s^*D^*K^*$}}(-g^{\varphi m}+\frac{k_2^{\varphi}k_2^{m}}{m_{D^*_s}})}{(k_1^2-m_{J/\psi}^2)(k_2^2-m_{D_s^*}^2)(k_3^2-m_{D^*}^2)}\nonumber\\
&&\varepsilon_{\mu \nu_1 \alpha_1 \beta_1} (i {k_2}_{n} g_{a m}-i {k_2}_{a} g_{n m}-i {k_3}_{m} g_{n a}+i {k_2}_{n} g_{\beta m}-i {p_1}_{ \beta} g_{n m}+i {p_1}_{ m} g_{n \beta})\Big\},\nonumber\\
{\mathcal M}_{{T_{\eta_c D^*}}}\!\!&=&\!\! \int \frac{d^4k_{\text{\tiny $3$}}}{(2\pi)^4}\Big(\frac{m_{B_c}^2-m_{\eta_c}^2}{k_2^2} {k_2}_{\rho}F_0(k_2^2)+(p+k_1-\frac{m_{B_c}^2-m_{\eta_c}^2}{k_2^2} k_2)_{\rho}{F}_1(k_2^2)\Big)\varepsilon^*_{\mu}(p_2)\varepsilon^{*n}(p_1)\nonumber\\
&&g_{\text{\tiny $T_{\eta_c D^*}$}}\mathcal{F}_2(k_3^2)(-g^{\beta \mu}+\frac{{k^\beta_3}{k^\mu_3}}{m_{D^*}^2})
\Big\{\frac{-g_{\text{\tiny $D_sD^*K^* $}}\varepsilon_{m n \alpha \beta}p^{m}_1 k_3^\alpha k_2^\rho i f_{D_s}}{(k_{\text{\tiny $1$}}^2-m_{\text{\tiny $\eta_c$}}^2)\ (k_{\text{\tiny $2$}}^2-m_{\text{\tiny $D_s$}}^2)\ (k_{\text{\tiny $3$}}^2-m_{\text{\tiny $D^*$}}^2)}({k_2}_{\beta} g_{n m}- {k_2}_{n} g_{\beta m}\nonumber\\&&- {k_2}_{n} g_{m \beta}+ {p_1}_{ \beta} g_{n m}- {p_1}_{ m} g_{n \beta})+\frac{i{m_{D_s^*}f_{D_s^*}g_{D_s^*D^*K^*}(-g^{\rho m}+\frac{k_2^{\rho}k_2^{m}}{m_{D^*_s}})}}{(k_{\text{\tiny $1$}}^2-m_{\text{\tiny $J/\psi$}}^2)\ (k_{\text{\tiny $2$}}^2-m_{\text{\tiny $D^*_s$}}^2)\ (k_{\text{\tiny $3$}}^2-m_{\text{\tiny $D^*$}}^2)} \Big\}.
\end{eqnarray}
In addition, to remove the ultraviolet divergence of amplitudes, we introduce the form factor $\mathcal{F}_2(k_3^2)$~\cite{Cheng:2004ru}
\begin{eqnarray}
 \mathcal{F}_2(k_3^2)=\Big(\frac{m_{D^*}^2-\Lambda_{D^*}^2}{k_3^2-\Lambda_{D^*}^2}\Big)^2,
\end{eqnarray}
where $m_{D^{{*}}}$ and $k_3$ denote the mass and momentum of the exchanged meson $D^{{*}}$ in the triangle diagrams. The cutoff parameter $\Lambda_{D^*}=m_{D^{{*}}}+\alpha_m$ (with $\alpha_m=\alpha \Lambda_{QCD} \approx$0.4 GeV)\cite{Liu:2024ziu} is used to suppress the contribution of hadrons at short distances.
\begin{figure}
	\includegraphics[width=0.99\columnwidth]{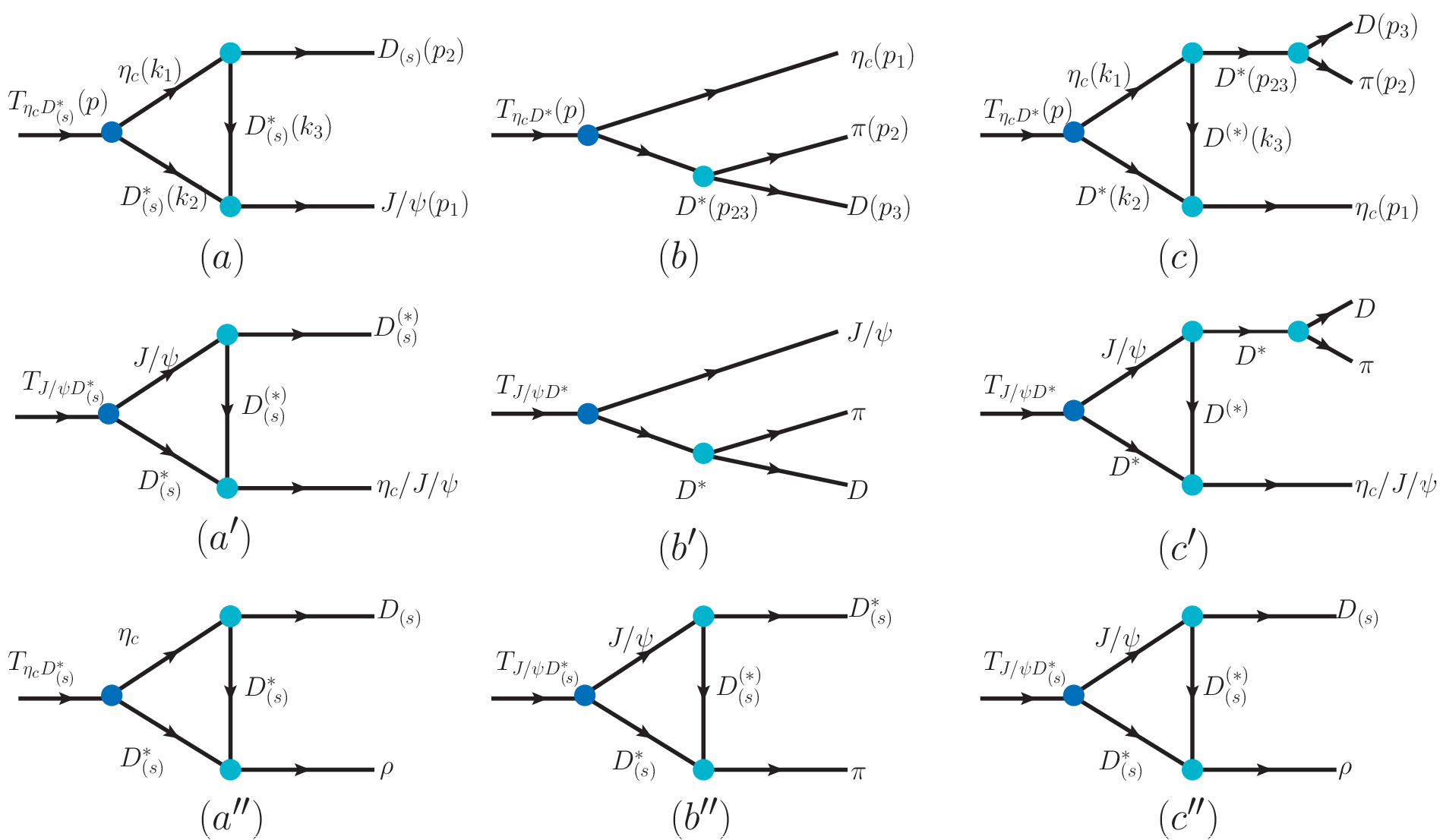}
	\caption{ Diagrams($a$,$a^\prime$,$a^{\prime \prime}$,$b^{\prime \prime}$,$c^{\prime \prime}$) represent two-body decays of the molecular states $T_{\eta_c D^*}$ and $T_{J/\psi D^*}$. Among them, the decay processes depicted in diagrams($a$,$a^\prime$)  are dominant, which are the primary focus of the paper, whereas the processes in diagrams($a^{\prime \prime}$,$b^{\prime \prime}$,$c^{\prime \prime}$) are suppressed.  Diagrams($b, b^\prime$) and ($c,c^\prime$) picture tree level and one-loop level three-body decays of the molecular states($T_{\eta_c D^*}$ and $T_{J/\psi D^*}$) respectively.}
\label{fig:Tcdecay}
\end{figure}
\subsubsection{Decay of the molecular states }
Within the effective Lagrangian approach, the decay processes of the molecular states $\eta_c D^*$ and $J/\psi D^*$ can be easily derived, their diagrams are presented in Fig.\,\ref{fig:Tcdecay}. In addition to the conventional two-body decays, we also investigate possible three-body decays, whose diagrams are shown in Fig.\,\ref{fig:Tcdecay}. In the work, we discuss the contribution of three-body decays arising at the one-loop level. The decay amplitudes $\mathcal M^{\prime}$ for the molecular states $T_{J/\psi D^*}$ and $T_{\eta_c D^*}$ are given as follows,
\begin{eqnarray}
&&\!\!\!\!\!\!\!\!{\mathcal M^{\prime}}^{a}_{T_{\eta_c D^*}\to{J/\psi} D}\nonumber\\
&&=\!\! \int \frac{d^4k_{\text{\tiny $3$}}}{(2\pi)^4}g_{\text{\tiny $\eta_cD^*D $}}g_{\text{\tiny $T_{\eta_c D^*}$}}g_{\text{\tiny $J/\psi D^*D^* $}}\frac{(-g^{m \mu}+\frac{k_3^m k_2^\mu}{m^2_{D^*}})(-g^{n \rho}+\frac{k_2^n k_2^\rho}{m_{D^*}^2}) \varepsilon_{\rho}(p) \varepsilon^{* a}(p_1)}{-(k_{\text{\tiny $1$}}^2-m_{\text{\tiny $\eta_c$}}^2)\ (k_{\text{\tiny $2$}}^2-m_{\text{\tiny $D^*$}}^2)\ (k_{\text{\tiny $3$}}^2-m_{\text{\tiny $D^*$}}^2)}\mathcal{F}_2(k_3^2) \nonumber\\
&& {(k_1+p_2)}_{\mu}(-i {k_2}_{n} g_{am}+i k_{3a} g_{m n}-i{k_2}_{m} g_{an}-i {k_2}_{n} g_{m a}+i {p_1}_{a} g_{n m}-i {p_1}_{ m} g_{n a}),
\nonumber\\
&&\!\!\!\!\!\!\!\!{\mathcal M^{\prime}}^{a^\prime}_{T_{J/\psi D^*}\to{J/\psi} D}\nonumber\\
&&=\!\! \int \frac{d^4k_{\text{\tiny $3$}}}{(2\pi)^4}g_{\text{\tiny $T_{J/\psi D^*}$}}\Big\{g_{\text{\tiny $D^*D^* J/\psi $}}g_{\text{\tiny $J/\psi D^*D $}}\frac{(-g^{\beta m}+\frac{k_3^\beta k_3^m}{m_{D^*}^2})(-g^{\xi n}+\frac{k_2^n k_2^{\xi}}{m_{D^*}^2})(-g^{\rho \nu}+\frac{k_1^\rho k_1^\nu}{m_{J/\psi}^{2}})}{(k_{\text{\tiny $1$}}^2-m_{\text{\tiny $J/\psi$}}^2)\ (k_{\text{\tiny $2$}}^2-m_{\text{\tiny $D^*$}}^2)\ (k_{\text{\tiny $3$}}^2-m_{\text{\tiny $D$}}^2)}\nonumber\\&&
(-{k_2}_{n} g_{a m}+{k_3}_{a}g_{m n}-k_{2 m} g_{a n}+{k_2}_{n} g_{m a}+{p_1}_{a} g_{n m}-{p_1}_{ m} g_{n a})\mathcal{F}_2(k_3^2)\nonumber\\
&&\varepsilon_{\mu \nu \alpha \beta} \varepsilon_{\gamma\sigma\rho\xi} \varepsilon^{\sigma}(p)\varepsilon^{* a}(p_1) k_1^\mu k_3^\alpha   p^\gamma  +g_{\text{\tiny $\eta_cD^*D $}}g_{\text{\tiny $J/\psi DD $}}\varepsilon^{* \nu} (p_1)\mathcal{F}_2(k_3^2)\nonumber\\&&
\frac{(-g^{\beta \xi}+\frac{k_2^\beta k_2^\xi}{m^2_{D^*}})(-g^{\mu \rho}+\frac{k_1^\mu k_1^\rho}{m_{J/\psi}^2})}{(k_{\text{\tiny $1$}}^2-m_{\text{\tiny $J/\psi$}}^2)\ (k_{\text{\tiny $2$}}^2-m_{\text{\tiny $D^*$}}^2)\ (k_{\text{\tiny $3$}}^2-m_{\text{\tiny $D$}}^2)} p_1^\mu k_2^\alpha \varepsilon^\sigma(p) \varepsilon_{\gamma \sigma \rho \xi} \varepsilon_{\mu \nu \alpha \beta}{(k_3-p_2)}_ {\mu} p^\gamma\Big\},\nonumber\\
&&\!\!\!\!\!\!\!\!{\mathcal M^{\prime}}^{a^\prime}_{T_{J/\psi D^*}\to{\eta_c D^*}}\nonumber\\&&=\!\! \int \frac{d^4k_{\text{\tiny $3$}}}{(2\pi)^4}g_{\text{\tiny $T_{J/\psi D^*}$}}\Big\{g_{\text{\tiny $\eta_cD^*D $}}g_{\text{\tiny $J/\psi DD $}}\frac{(-g_{\mu \rho}+\frac{k_2^\mu k_2^p}{m_{D^*}^2})(-i p^\sigma) \varepsilon_{\sigma \nu \alpha \rho}(-g^{\alpha \gamma}+\frac{k_1^\alpha k_1^\gamma}{m_{J/\psi}^2})}{(k_{\text{\tiny $1$}}^2-m_{\text{\tiny $J/\psi$}}^2)\ (k_{\text{\tiny $2$}}^2-m_{\text{\tiny $D^*$}}^2)\ (k_{\text{\tiny $3$}}^2-m_{\text{\tiny $D$}}^2)}\mathcal{F}_2(k_3^2) \nonumber\\
&& {(p_2+k_3)}_{\mu} k_1^m p_1^n \varepsilon^\nu(p) \varepsilon_{m A x} \varepsilon_{n \xi m \gamma} \varepsilon^{* \xi}(p_2)+  g_{\text{\tiny $\eta_cD^*D^* $}}g_{\text{\tiny $J/\psi D^*D $}}\varepsilon^{* n}(p_1)\mathcal{F}_2(k_3^2)\nonumber\\
&&\frac{(-g_{a \xi}+ \frac{{k_3}_{a}{k_3}_{\xi}}{m^2_{D^*}})(-g_{\sigma \beta}+\frac{{k_2}_{\beta} {k_2}_{\sigma}}{m^2_{D^*}})(-g_{\rho m}+\frac{{k_1}_{\rho}k_{1 m}}{m_{J/\psi}^2}) } {i(k_{\text{\tiny $1$}}^2-m_{\text{\tiny $J/\psi$}}^2)\ (k_{\text{\tiny $2$}}^2-m_{\text{\tiny $D^*$}}^2)\ (k_{\text{\tiny $3$}}^2-m_{\text{\tiny $D$}}^2)} \varepsilon^{\mu \nu \alpha \beta} \varepsilon^{\gamma \xi \rho \sigma} p_\gamma  {k_3}_{\gamma} {k_2}_{\alpha}\nonumber\\
&&(- {k_1}_{n} g^{a m}+ k_1^a g_n^m+ k_3^m g_n^a- {k_2}_{n} g^{a m}+ p_2^a g_n^m- p_2^m g_n^a)\Big\} ,
\nonumber\\
&&\!\!\!\!\!\!\!\!{\mathcal M^{\prime}}^{b^{\prime}+c^{\prime}}_{T_{\eta_c D^*}\to \eta_c D\pi}\nonumber\\
&&=g_{\text{\tiny $T_{\eta_c D^*}$}}g_{\text{\tiny $ D^* D\pi$}} \frac{\varepsilon^{\mu}(p)\left(-g_{\mu \nu}+\frac{{p_{23}}_\mu {p_{23}}_{\nu}}{m^2_{D^*}}\right)\left(p_2-p_3\right)^\nu}{\left(p_{23}^2-m_{D^*}^2\right)}
+\!\! \int \frac{d^4k_{\text{\tiny $3$}}}{(2\pi)^4}g_{\text{\tiny $T_{\eta_c D^*}$}}\nonumber\\
&&\Big\{g^2_{\text{\tiny $\eta_c D^*D^*$}}g_{\text{\tiny $D^* D \pi $}}\varepsilon_n(p)\mathcal{F}_2(k_3^2)\frac{ (-g^{\mu \nu}+\frac{p_{23}^\nu p_{23}^\mu}{m^2_{D^*}})(-g^{m n}+\frac{k_2^m k_2^n}{m_{D^*}^2})}{i(k_{\text{\tiny $1$}}^2-m_{\text{\tiny $\eta_c$}}^2)\ (k_{\text{\tiny $2$}}^2-m_{\text{\tiny $D^*$}}^2)\ (k_{\text{\tiny $3$}}^2-m_{\text{\tiny $D$}}^2)} \nonumber\\
&& {( p_{2 }- p_{3})}_{\nu}{(k_1-k_3 )}_{\mu}){(k_{3 }-p_{1 })}_m +
g^2_{\text{\tiny $\eta_c D^*D$}}g_{\text{\tiny $D^* D \pi $}}\varepsilon_{\sigma}(p)\varepsilon_{\xi n m \gamma} k_2^{\xi} k_3^m\nonumber\\
&&(p_2-p_3)_{\rho}\mathcal{F}_2(k_3^2)\varepsilon_{\mu\nu\alpha\beta}k_3^{\mu}p_{23}^{\alpha}\frac{ (-g^{\sigma n}+\frac{k_{2}^{\sigma}k_{2}^{n}}{m_{D^{*}}^2})(-g^{\nu \gamma}+\frac{k_{3}^{\nu} k_{3}^{\gamma}}{m_{D^{*}}^2}) (-g^{\beta \rho}+\frac{p_{23}^{\beta} p_{23}^{\rho}}{m_{D^{*}}^2})}{(k_{\text{\tiny $1$}}^2-m_{\text{\tiny $\eta_c $}}^2)\ (k_{\text{\tiny $2$}}^2-m_{\text{\tiny $D^*$}}^2)\ (k_{\text{\tiny $3$}}^2-m_{\text{\tiny $D^*$}}^2)}\Big\}\,,\nonumber\\
&&\!\!\!\!\!\!\!\!{\mathcal M^{\prime}}^{b+c}_{T_{J/\psi D^*}\to J/\psi D\pi}\nonumber\\
&&=g_{\text{\tiny $T_{J/\psi D^*}$}}g_{\text{\tiny $ D^*D \pi $}}\frac{\varepsilon^{\mu\nu\alpha\beta}\varepsilon_{\nu}(p)\varepsilon^{*}_{\alpha}(p_1)(-g_{\beta m}+\frac{{P_23}_{\beta} {P_23}_{m} }{m^2_{D^*}})p_\mu {(p_2-p_3)}^m}{\left(p_{23}^2-m_{D^*}^2\right)}\nonumber\\
&+&\!\! \int \frac{d^4k_{\text{\tiny $3$}}}{(2\pi)^4}g_{\text{\tiny $T_{J/\psi D^*}$}}\Big\{g_{\text{\tiny $J/\psi D^*D^*$}}g_{\text{\tiny $\eta_c D^*D^*$}}g_{\text{\tiny $D^* D \pi $}} \varepsilon_{\alpha\beta \rho \sigma} p^{\alpha} \varepsilon ^ { \beta } ( p ) \varepsilon ^ { * a }({p_1}) (-g^{\rho \mu}+ \frac{k_1^{\rho}k_1^{\mu}}{m_{J/\psi}^2})\nonumber\\
&&(p_2-p_3)_{ \nu}\mathcal{F}_2(k_3^2)(-g^{\sigma n}+\frac{k_2^{\sigma} k_2^{n}}{m_{D^{*}}^2})\frac{
(-g^{\gamma \nu}+\frac{p_{23}^{\gamma}p_{23}^{\nu}}{m_{D^{*}}^2})
(-g^{m \xi}+\frac{k_{3}^{m}k_{3}^{\xi}}{m_{D^{*}}^2})}
{i(k_1^2-m_{J/\psi}^2)(k_2^2-m_{D^{*}}^2)(k_3^2-m_{D^{*}}^2)} \nonumber\\
& & ( {k_2}_{n} g_{a m}- {k_3}_{a}g_{m n}+{k_2}_{m} g_{a n}+ {k_2}_{n} g_{m a} - {p_1}_{a}g_{n m}+ {p_1}_{m} g_{n a})((p _1+k_3)_{\gamma} g_{\xi \mu}\nonumber\\
& &-{(p_1+k_3)}_{\xi} g_{\gamma \mu}-{k_3}_{\mu} g_{\xi \gamma}+ {k_3}_{\gamma} g_{\xi \mu}- {p_1}_{\xi} g_{\gamma \mu}-{p_1}_{\mu} g_{\gamma \xi})+g_{\text{\tiny $J/\psi D^*D^*$}}g_{\text{\tiny $\eta_c D^*D^*$}}\nonumber\\
&&g_{\text{\tiny $D^* D \pi $}}  \varepsilon_{\alpha\beta \rho \sigma}\varepsilon_{n a m \mu} \varepsilon_{\zeta \xi r \gamma } p^{\alpha} \varepsilon ^ { \beta } ( p ) \varepsilon ^ { * a } ( p_{ 1 } )p^n_1 k^m_2\varepsilon ^ { * a } ( p_{ 1 } ) k^{\zeta}_1 p^r_{23} {(p_2-p_3)}_{\nu}\mathcal{F}^2(k_3^2)\nonumber\\
&&(-g^{\rho \xi}+ \frac{k_1^{\rho}k_1^{\xi}}{m_{J/\psi}^2})\frac{
(-g^{\sigma \mu}+\frac{k_2^{\sigma} k_2^{\nu}}{m_{D^{*}}^2})(-g^{\gamma \nu}+\frac{p_{23}^{\gamma}p_{23}^{\nu}}{m_{D^{*}}^2})
}
{(k_1^2-m_{J/\psi}^2)(k_2^2-m_{D^{*}}^2)(k_3^2-m_{D}^2)}\Big\},\nonumber\\
&&\!\!\!\!\!\!\!\!{\mathcal M^{\prime}}^{c^{\prime}}_{T_{J/\psi D^*}\to \eta_c D\pi}\nonumber\\
&&=\!\! \int \frac{d^4k_{\text{\tiny $3$}}}{(2\pi)^4}g_{\text{\tiny $T_{J/\psi D^*}$}}g_{\text{\tiny $D^* D \pi $}}\Big\{g_{\text{\tiny $J/\psi D^*D^*$}}g_{\text{\tiny $\eta_c D^*D^*$}}\frac{i\varepsilon_{\xi \zeta \rho\sigma}\varepsilon^{\xi}(p)p^{\zeta}(-g^{\rho m}+\frac{k_1^{\rho}k_1^m}{m_{J/\psi}^2})}{(k_{\text{\tiny $1$}}^2-m_{\text{\tiny $J/\psi$}}^2)\ (k_{\text{\tiny $2$}}^2-m_{\text{\tiny $D^*$}}^2)\ (k_{\text{\tiny $3$}}^2-m_{\text{\tiny $D^*$}}^2)}\nonumber\\
&&(-g^{\sigma \beta}+\frac{k_2^{\sigma}k_2^{\beta}}{m_{D^{*}}^2})\mathcal{F}_2(k_3^2) ((p_1+k_3)_{n} g_{am}-(p_1+k_3)_{a} g_{n m}
-{k_3}_{m} g_{n a}+ {k_2}_{n} g_{a m}\nonumber\\
&&- {p_1}_{a} g_{n m}-{p_1}_{ m} g_{n a})\varepsilon_{\mu\nu\alpha\beta}k_3^{\mu}k_2^{\alpha}(-g^{\nu a}+\frac{k_3^{\nu}k_3^{a}}{m_{D^{*}}^2})(-g^{n \gamma }+\frac{p_{23}^{n}p_{23}^{\gamma}}{m_{D^{*}}^2})(p_{23}-p_{3})_{\gamma})\nonumber\\
&&+g_{\text{\tiny $J/\psi DD^*$}}g_{\text{\tiny $\eta_c DD^*$}}\varepsilon_{\rho\nu\alpha\beta}p_{23}^{\rho}k_1^{\alpha} \mathcal{F}_2(k_3^2){(p_{2}-p_{3})}_{\gamma} (p_1+k_3)_\mu  \varepsilon_{m n \xi \sigma} \varepsilon^n(p) p^m \nonumber\\
&& \frac{(-g^{\gamma\nu}+\frac{p_{23}^{\gamma}p_{23}^{\nu}}{m_{D^{*}}^2})(-g^{\beta \xi}+\frac{k_1^{\beta}k_1^{\xi}}{m_{J/\psi}^2})(-g^{\mu \sigma}+\frac{k_2^\mu k_2^\sigma}{m_{D^{*}}^2})}{i(k_{\text{\tiny $1$}}^2-m_{\text{\tiny $J/\psi$}}^2)\ (k_{\text{\tiny $2$}}^2-m_{\text{\tiny $D^*$}}^2)\ (k_{\text{\tiny $3$}}^2-m_{\text{\tiny $D$}}^2)}\Big\}.\nonumber\\
\end{eqnarray}
The momentum conventions are drawn in Fig.\,\ref{fig:Tcdecay}, the momentum $p_{23} = p_2 + p_3$ and $k_3$ is the momentum of the exchanged particle. The partial decay width for the three-body process $T_{J/\psi(/\eta_c) D^*} \to J/\psi(/\eta_c) D \pi$ can be expressed as~\cite{ParticleDataGroup:2024cfk}:
\begin{eqnarray}
d \Gamma =\frac{1}{(2 \pi)^3} \frac{1}{32 m^3_{\text{\tiny $T$}}} \overline{|\mathcal{M}^{\prime}|^2} d m_{12}^2 d m_{23}^2.
\end{eqnarray}
$\mathcal{M}^{\prime}$ denotes the decay amplitudes, $m_{12}$ and $m_{23}$ represent the invariant masses of final $J/\psi \pi$(or $\eta_c \pi$) and $D \pi$ system, respectively, defined as $
m^2_{12} = (p_1+p_2)^2$ and
$m^2_{23} = (p_2+p_3)^2$. The limits for the outer integration variable \( m_{23}^2 \) and  \( m_{12}^2 \) are given by,
\begin{eqnarray}
&&(m_{23}^2)_{\min} = (m_2 + m_3)^2, \quad (m_{23}^2)_{\max} = (M - m_1)^2,\notag\\
&&(m_{12}^2)_{\max} = (E_1^* + E_2^*)^2 - \left( \sqrt{{E_1^*}^2 - m_1^2} - \sqrt{{E_2^*}^2 - m_2^2} \right)^2,\notag\\
&&(m_{12}^2)_{\min} = (E_1^* + E_2^*)^2 - \left( \sqrt{{E_1^*}^2 - m_1^2} + \sqrt{{E_2^*}^2 - m_2^2} \right)^2,
\end{eqnarray}
where $
E_1^* = ({M^2 - m_{23}^2 - m_1^2})/{2m_{23}}$ 
 and $E_2^* =({m_{23}^2 - m_3^2 + m_2^2})/{2m_{23}}$, 
with \( E_1^* \) and \( E_2^* \) are the energies of $J/\psi$ (or $\eta_c$) and $\pi$ respectively in the rest frame of $m_{23}$.
\section{Numerical analysis}
\label{Numerical analysis and discuss}
The form factors in Eq.(\ref{eq:formfactor1}) and Eq.(\ref{eq:formfactor2}) can be parameterized on the physical region,
\begin{eqnarray*}
f(q^2)=\alpha_1 +\alpha_2 q^2+\frac{\alpha_3 q^4}{m_{B_c}^2-q^2} \,,
\end{eqnarray*}
where the fitted parameters $\alpha_{1,2,3}$ are taken from reference~\cite{Yao:2021pyf}. The decay constants, coupling constants and Wilson coefficient $a_1$ are collected into Tab.\,\ref{tab:input1}. The coupling constants $g_{\text{\tiny $T_{\text{\tiny $\eta_c D^*$}}$}}$ and $g_{\text{\tiny $T_{\text{\tiny $J/\psi D^*$}}$}}$ between the  molecular states $cc\bar c\bar q$ state and their constituent can be obtained by Eq.\,(\ref{eq3}).  The couplings does not vary significantly with the scale parameter $\Lambda_T$. We take $\Lambda_T$ from 0.8 to 1.2 GeV, the variation in the strong coupling constant is less than $6\%$. The three-meson strong couplings constants are derived from the SU(4) flavor symmetry analysis. It should be emphasized that incorporating $\text{SU}(4)$ flavor symmetry breaking effects into our calculation is challenging, as this symmetry serves as the foundation of the hadronic effective Lagrangian we employ. Nevertheless, to assess how potential uncertainties in these parameters, we varied the coupling constants by $\pm 10\%$ in the numerical analysis.

\begin{figure}
	\includegraphics[width=0.70\columnwidth]{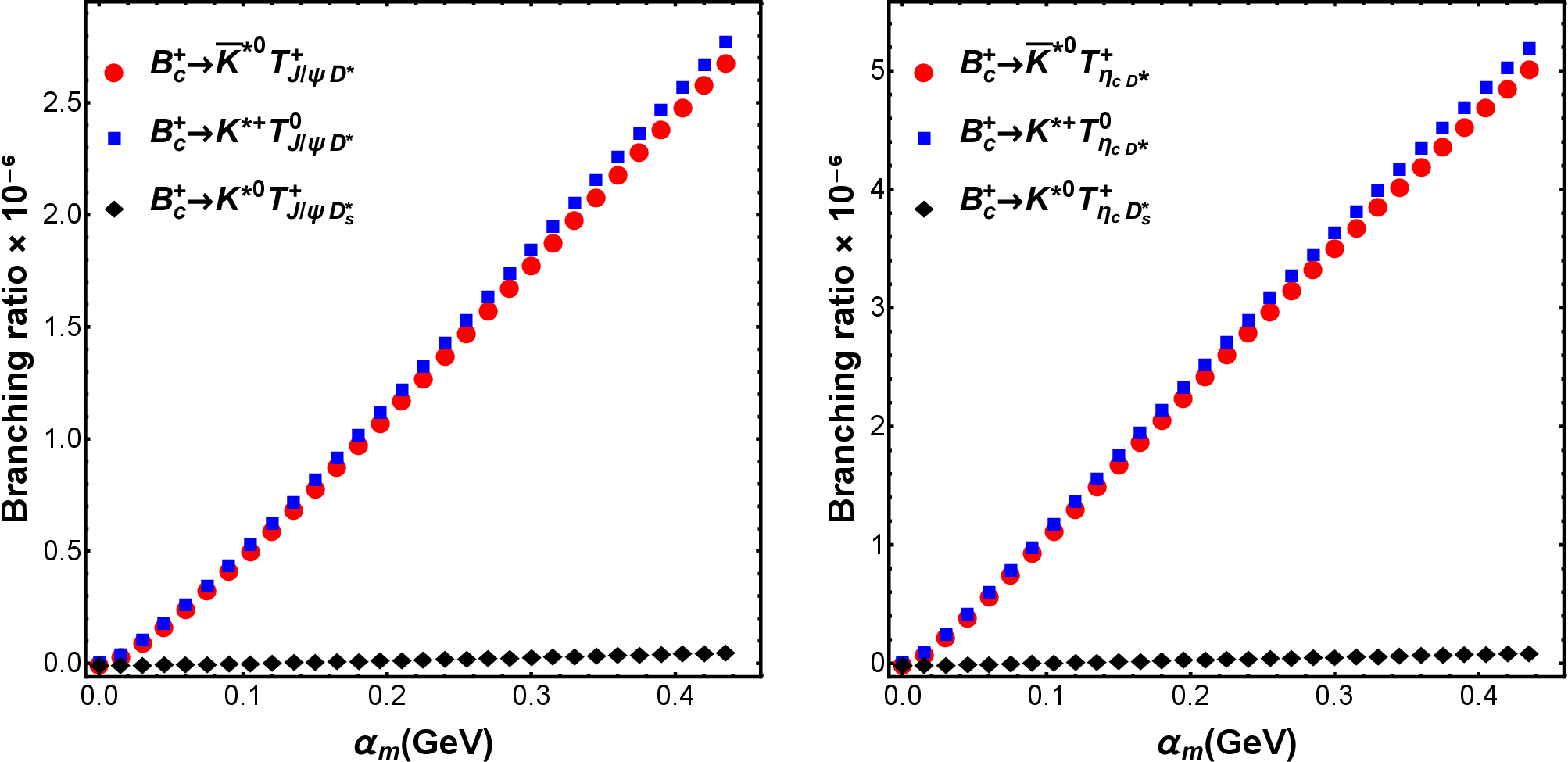}
	\caption{ In the binding energies $\varepsilon = 5\ \text{MeV}$,
        the branching ratios of $B^+_c \to K^* T_{cc\bar c \bar q}$ vary with the scale parameter $\alpha_m$(GeV).
        The panels correspond to $J/\psi D^*$ molecular configuration (left),
       and $\eta_c D^*$ molecular configuration (right).}
\label{fig:production}
\end{figure}

\begin{figure}
	\includegraphics[width=0.99\columnwidth]{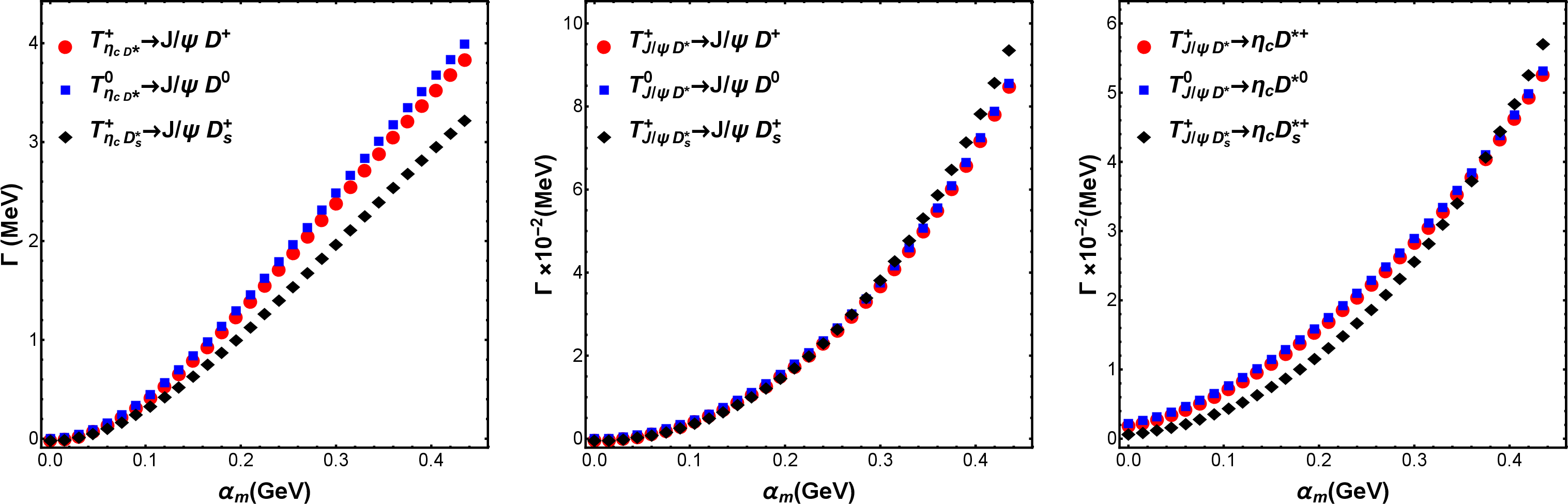}
	\caption{ In the binding energies $\varepsilon = 5\ \text{MeV}$,
        the two-body decay widths of $T_{cc\bar c \bar q}$ vary with the scale parameter $\alpha_m$(GeV).
        The panels correspond to binding energies $T_{ \eta_c D^*}  \to     J/\psi D$ (left),
        $T_{ J/\psi D^*}  \to     J/\psi D$ (middle), and $T_{ J/\psi D^*}  \to     \eta_c D^*$ (right).}
\label{fig:alphaTcdecay}
\end{figure}

{In the schemes of molecular states $T_{J/\psi D^*}$ and $T_{\eta_c D^*}$, the production branching ratios from $B_c$ meson are listed in Tab.~\ref{tab:pro}, where we take cutoff parameter $\alpha_m=0.4 $GeV and binding energy $\epsilon= 5, 10, 20$ MeV. Besides, the decay widths are obtained in Tab.~\ref{tab:dec}, in which the results not provided are mainly forbidden by the effect of phase space. For completeness, the dependence of the production branching ratios and the two-body decay widths of the four-quark states on the cutoff parameter $\alpha_m$ are shown in Fig.\ref{fig:production} and Fig.\ref{fig:alphaTcdecay}, respectively, with $\alpha_m$ varying in the range of $100-450$ MeV~\cite{Cheng:2004ru}. Note that the widths of three-body decays are generally small, especially for processes with only loop-diagram contributions, and they depend very weakly on $\alpha_m$, though a slight increasing trend is observed as $\alpha_m$ increases.}
\begin{itemize}
\item
In general, the production branching ratios for $T_{\eta_c D^*}$ molecular configuration are larger than those with molecule $T_{J/\psi D^*}$. Specifically, the branching ratios of two Cabibbo-allowed processes $B_c^+   \to \overline  K^{*0}   T_{\eta_c D^* }^{+}$ and $B_c^+ \to   K^{*+}   T_{\eta_c D^*}^{0}$, can reach the order of $10^{-4}$.
\begin{eqnarray}
&&Br(B_c^+   \to \overline  K^{*0}   T_{\eta_c D^*}^{+})=1.127\times10^{-4},\,\,\,\,  Br(B_c^+ \to   K^{*+}   T_{\eta_c D^*}^{0})=1.162\times10^{-4},\nonumber\\
&&Br(B_c^+   \to \overline  K^{*0}   T_{J/\psi D^*}^{+})=8.160\times10^{-6},\,\,\,\,  Br(B_c^+ \to   K^{*+}   T_{J/\psi D^*}^{0})=8.400\times10^{-6}.
\end{eqnarray}
\end{itemize}
\begin{itemize}
\item
We define two ratios $R_1$ and $R_2$ for the production processes,
\begin{eqnarray}
R_1=\frac{{\Gamma(B_c^+\to K^{*+}  T_{J/\psi D^*}^{0})}}{\Gamma(B_c^+\to \overline{K}^{*0} T_{J/\psi D_s^*}^{+})}\approx 42,\quad R_2=\frac{\Gamma(B_c^+\to \overline{K}^{*0} T_{\eta_c D^*}^{+})}{\Gamma(B_c^+\to \overline{K}^{*0} T_{\eta_c D_s^*}^{+})}\approx 50.7,
\end{eqnarray}
which are different from SU(3) symmetric prediction $R_{1,2}=|V_{cs}|^2/|V_{cd}|^2 \approx 19.9$ in Eq.\,(\ref{eqn8}). These indicate that the two processes encounter significant SU(3) flavor breaking effect. Phase space effects and the dynamics of loop diagrams are likely the primary mechanisms responsible for these deviations. In loop and phase space integrations, the propagator of the exchanged particle $\mathcal{S}_{ex}(k_3)$ in the triangle diagram and the strong form factors $\mathcal{F}_2(k_3^2)$ at the vertices give rise to final-state dynamical effects that deviate significantly from the SU(3) symmetry analysis, as a result of the differences in the masses, decay constants, and coupling constants of the $D$ and $D_s$ mesons.
\end{itemize}
\begin{itemize}
\item
It should be noted that the leading order contribution of some three-body processes comes from tree hadronic diagrams, such as processes $T^+_{\eta_c D^*}   \to  D^0 \eta_c \pi^+$ and $T^+_{J/\psi D^*}   \to    D^0 J/\psi \pi^+$, the calculations begin with triangle diagram as Fig.\,\ref{fig:Tcdecay}.($b$,$b^\prime$).
\end{itemize}
\begin{itemize}
\item
In addition to the cutoff parameter $\alpha_m$ and the binging energy $\epsilon$, the uncertainties in the production and decay processes primarily originate from the strong coupling constants associated with the $T_{\eta_c D^*}$ and  $T_{J/\Psi D^*}$ states and the three-meson interactions. Specifically, variations in the former ones lead to a $12\%$ change in the production branching fraction and a $8\%$ change in the decay width, respectively. When the $\pm10\%$ variation in the three-meson coupling constants, arising from SU(4) flavor symmetry breaking, is further considered, the production branching fraction and decay width exhibit changes of $20\%$ and $32\%$, respectively. Combining both sources of uncertainty, the total production branching fraction and decay width is modified by $24\%$ and $26\%$ finally.
\end{itemize}
\begin{itemize}
\item
Summing up all possible two-body and three-body processes, we derive total decay width of the molecular states $T_{\eta_c D^*}$ and $T_{J/\Psi D^*}$ with $J^P=1^+$, at the cutoff parameter $\alpha=0.4$ GeV, and the binding energy $\epsilon=5$ MeV,
\begin{eqnarray}
\Gamma(T^+_{\eta_c D^*})=6.8 \text{MeV},\quad
\Gamma(T^+_{J/\psi D^*})=0.16 \text{MeV},
\end{eqnarray}
Given the existing studies based on different methodologies and internal structures model, we perform a direct comparison. Our results are consistent with those calculated using the Gaussian Expansion Method(GEM)~\cite{Yang:2024nyc}, in which the authors predict a narrow resonance in a full coupled-channel calculation with a total width of approximately $\Gamma(cc\bar c\bar q)=3.0$ MeV.

As a comparison, the compact tetraquark scheme within constituent quark model~\cite{ Li:2025fmf} gives a relatively large decay widths, $\Gamma(cc\bar c\bar q)=240.8$ MeV, whose significant contributions coming from decay modes $J/\psi D^*$, $J/\psi D$, and $\eta_c D^*$. The relative decay widths are about $\Gamma(J/\Psi D^*):\Gamma(\eta_c D^*):\Gamma(J/\Psi D)=0.004:1.3:1$~\cite{Li:2025fmf,Weng:2021ngd} for the lowest $J^P=1^+$ compact tetraquark states.
\end{itemize}
\begin{itemize}
\item
For completeness, we further analyze their decays into final states containing light mesons, such as $D\rho$ and $D^*\pi$. These decay channels involve the annihilation of the $c\overline{c}$ component and the creation of light quark pairs, and are subject to both OZI and $\alpha_s$ order suppressions, in addition, it involves the exchange of hard gluons and are therefore subject to  dynamical suppression due to the heavy charmed quark mass. Nevertheless, we employ the effective Lagrangian approach to compute these processes. The triangle diagrams are shown in Fig~\ref{fig:Tcdecay}.($a^{\prime \prime}$,$b^{\prime \prime}$,$c^{\prime \prime}$). For $T^+_{\eta_c D^*}$ and $T^+_{J/\psi D^*}$ configurations, the decay widths with $\epsilon=20$MeV are
\begin{eqnarray}
&\Gamma(T^+_{\eta_c D^*} \to \rho^0 D^+) = 0.537\ \text{MeV},
&\Gamma(T^+_{\eta_c D^*} \to D^{*+}\pi^0) = 0.290\ \text{MeV},\\
&\Gamma(T_{J/\psi D^*}^+ \to \rho^0 D^+) = 0.0038\ \text{MeV},
&\Gamma(T_{J/\psi D^*}^+ \to D^{*+}\pi^0) = 0.0038\ \text{MeV}.
\end{eqnarray}
They are about one or two orders of magnitude smaller than the two-body decay widths listed in Table~\ref{tab:dec}. Therefore, such $c\bar c$ annihilation decay processes do not constitute the dominant contribution, for convenience, we do not provide a detailed discussion of such decay processes involving in the present paper.
\end{itemize}
\begin{table}
\caption{The input parameters including decay constants, coupling constants and  Wilson coefficient.~\cite{Shen:2019evi,Yalikun:2021dpk,Xie:2022hhv}}
\label{tab:input1}
\begin{tabular}{lllllll}\hline\hline
&$f_{D}=0.200$GeV,\quad $f_{D_s}=0.241$GeV,\quad$f_{D^*}=0.200$GeV,\quad $f_{D^*_s}=0.241$GeV,\quad $a_{1}=1.07$,\\
&$a_2=-0.017$,\quad$g_{K^*D^*D}=-7.07$,\quad $g_{D^*D^*J/\psi}=2.298$,\quad $g_{D^*D^*K^*}=-2.298$,\\
&$g_{D^{*+}D^0\pi^+}=16.818$,\quad $g_{D^{*0}D^0\pi^0}=11.688$,\quad $g_{\eta_c DD^*}=-4.271$,\quad $g_{D^*D^*\eta_c}=-7.07$,\\
&$g_{J/\psi D^*D}=-7.07$,\quad $g_{T_{J/\psi D^*}}=1.223^{+0.045}_{-0.024}$,\quad $g_{T_{\eta_c D^*}}=16.065^{+0.935}_{-0.605}$
 \\\hline\hline
 \end{tabular}
\end{table}
\begin{table}
  \centering
\caption{ The production branching ratio ($\times 10^{-5}$) of the  molecular states $cc\bar c\bar q$ with binding energy $\epsilon=5,10,20$ MeV.}\label{tab:pro}
\renewcommand{\arraystretch}{0.8} \tabcolsep2pt%
       \begin{tabular}{ccccc|cccccc}\hline\hline
      \multicolumn{2}{c}{ \multirow{2}*{$\eta_c D^*$}}
   &  \multicolumn{3}{c|} { branching ratio ($\times 10^{-5}$)} & \multicolumn{2}{c}{\multirow{2}*{$J/\psi D^*$}}
   &  \multicolumn{3}{c} { branching ratio ($\times 10^{-5}$)}  \\ \cline{3-5}\cline{8-10}
    &&  $\epsilon=5$&\quad  $\epsilon=10$\quad & $\epsilon=20$ & &&  $\epsilon=5$&\quad $\epsilon=10$\quad &$\epsilon=20$   \\\hline
  $B_c^+$&$ \to  \overline  K^{*0}   T_{\eta_c D^*}^{+}$ &4.649 &$6.198$ &11.272&$B_c^+$&$   \to \overline  K^{*0}   T_{J/\psi D^*}^{+}$  &0.245 &$0.422$ &0.816\\
&$\to  K^{*+}   T_{\eta_c D^*}^{0}$ &4.798 &$6.393$ &11.621&&$ \to  K^{*+}    T_{J/\psi D^*}^{0}$  &0.253 &$0.435$ &0.840\\
&$\to   K^{*0}   T_{\eta_c D_s^*}^{+}$ &0.093 &$0.125$ &0.229&&$  \to       K^{*0}   T_{J/\psi D_s^*}^{+}$  &0.005 &$0.009$ &0.020
\\\hline\hline
    \end{tabular}
\end{table}
\begin{table}
  \centering
 \caption{The partial decay widths (in units of $MeV$) for the two-body and three-body decays of four-quark molecular states $T_{J/\psi D^*}$ and $T_{\eta_c D^*}$.}\label{tab:dec}
\renewcommand{\arraystretch}{1.3} \tabcolsep0.5pt%
       \begin{tabular}{ccccc|cccccc}\\\hline\hline
       \multicolumn{2}{c}{\multirow{2}*{process}}	
 &    \multicolumn{3}{c|}{ widths $/MeV$} &\multicolumn{2}{c}{\multirow{2}*{process}} &    \multicolumn{3}{c}{ widths  $/MeV$}  \\ \cline{3-5}\cline{8-10}
     && $\epsilon=5$ & $\epsilon=10$ &  $\epsilon=20$ & &&  $\epsilon=5$ & $\epsilon=10$ & $\epsilon=20$   \\\hline
$T^+_{\eta_c D^*} $&$  \to     J/\psi D^+$  &3.491 &4.093&$4.385$ &$T^0_{\eta_c D^*} $&$ \to   J/\psi D^0 $   &3.621&4.477 &$4.583$ \\
&$ \to     D^+  \eta_c \pi^0$  &0.998 &$-$ &$-$&&$ \to     D^0 \eta_c \pi^0$  &0.964 &$-$ &$-$\\
&$\to     D^0  \eta_c \pi^+$  &2.290 &$-$ &$-$&$T^+_{\eta_c D_s^*}$ &$ \to   J/\psi D^+_s $   &2.293 &$3.757$ &3.854\\\hline
$T^+_{J/\psi D^*}$&$   \to     J/\psi D^+$  &0.070 &$0.122$ &0.238&$T^0_{J/\psi D^*}   $&$\to     J/\psi D^0$  &0.070 &$0.123$ &0.239&\\
&$\to  \eta_c  {D^*}^+ $  &0.045 &$0.076$ &0.141&&$ \to   \eta_c {D^*}^0 $  &0.045&$0.077$ &0.142&\\
&$\to   D^+   J/\psi \pi^0$  &$0.015$ &$-$ &$-$& &$ \to   D^0  J/\psi \pi^0$  &$0.019$ &$-$ &$-$\\
&$ \to    D^0  J/\psi \pi^+$  &0.025 &$-$ &$-$&&$ \to   D^0 \eta_c \pi^0$  &$1.8\times10^{-4}$ &$3.2\times10^{-4}$ &$6.5\times10^{-4}$\\
&$\to    D^+ \eta_c \pi^0$  &$1.9\times10^{-4}$ &$3.5\times10^{-4}$ &$7.0\times10^{-4}$&$T^+_{J/\psi D_s^*}$&$   \to     J/\psi D^+_s$  &0.076 &$0.136$ &0.267\\
&$\to   D^0  \eta_c \pi^+$  &$2.4\times10^{-4}$&$4.3\times10^{-4}$ &$8.6\times10^{-4}$&&$\to     \eta_c {D_s^*}^+$  &0.047 &$0.083$ &0.163\\\hline
\hline
    \end{tabular}
\end{table}

\section{Conclusions}
In this work, we discuss the production of molecular states $T_{J/\psi D^*}$ and $T_{\eta_c D^*}$ through $B_c$ meson decays, as well as their two-body and three-body strong decays. Starting from SU(3) symmetric phenomenological analysis, we primarily choose several golden channels for the production and decay of these four-quark molecular states. Immediately after using the effective Lagrangian approach, the exhaustive decay widths and production branching ratios are shown up, based on the effective Lagrangian approach. We find that the production branching ratios increase with both the cutoff parameter $\alpha_m$ and the binding energy $\epsilon$. For the $T_{\eta_c D^*}$ configuration, the branching ratios can reach the order of $10^{-4}$. Furthermore,  it shows that the decay widths are not significant, which both at order of MeV regarding to strong decay of the two molecular configurations $T_{\eta_c D^*}$ and $T_{J/\psi D^*}$.
\begin{acknowledgments}
We thank Prof. LiSheng Geng for the meaningful discussion remarks and useful comments. This work is supported by NSFC under grants No. 12005294.
\end{acknowledgments}
\appendix
\section{}\label{sec:production}
The coupling constants ${g^2_{T_{\text{\tiny${\eta_c D^*}$}}}}$ and ${g^2_{T_{\text{\tiny${J/\psi D^*}$}}}}$ can be deduced as,
	\begin{equation}
\begin{aligned}
\frac{1}{{g^2_{T_{\text{\tiny${\eta_c D^*}$}}}} } =&3 \int_{0}^{1}d\alpha\int_{0}^{\infty}d\beta\frac{\beta}{16\pi^{2}} \frac{\partial^2}{\partial p^2}  \Big( \frac{-3\Phi^2(\Delta)}{(1+\beta)^2}+\frac{\Lambda_T^{2}\Phi^2(\Delta)}{2{m^2_{\text{\tiny$D^*$}}}}\Big(\frac{2}{(-1-\beta)^3}+\frac{1}{(1+\beta)^4}\times\frac{p^{2}}{2\Lambda_T^{2}}\\&\times(2(1-\alpha)\beta-\frac{-2{m_{\text{\tiny$D^*$}}}}{{m_{\text{\tiny$\eta_c$}}}+{m_{\text{\tiny$D^*$}}}})^2p^2\Big) -\frac{\Lambda_T^4\Phi^2(\Delta)}{4m_{T_{\text{\tiny$\eta_c D^*$}}}^2{m_{\text{\tiny$D^*$}}}^2}\Big(\frac{1}{(1+\beta)^4}\frac{p^4}{4\Lambda_T^4}\times(4(\alpha-1)\beta \nonumber\\
&+2\frac{-2m_{\text{\tiny$D^*$}}}{m_{\eta_c}+m_{\text{\tiny$D^*$}}})^2+\frac{1}{(-1-\beta)^3}\times\frac{p^2}{\Lambda_T^2}\Big)\Big),
\end{aligned}
\end{equation}
\begin{equation}
\begin{aligned}
\frac{1}{g^2_{T_{\text{\tiny${J/\psi D^*}$}}}} &=3  \int_{0}^{1}d\alpha\int_{0}^{\infty}d\beta\frac{\beta}{16\pi^{2}}\frac{\partial^2}{\partial p^2}\Big( \frac{-\Lambda_T^4\Delta_5}{2{m^2_{\text{\tiny$J/\psi$}}}}-\frac{6p^2}{(-1-\beta)^2} \Phi^2(\Delta)+\frac{\Lambda_T^2p^2\Delta_1 }{m^2_{\text{\tiny$J/\psi$}}} +\frac{p^2}{{m^2_{\text{\tiny${{ D^*}}$}}}{m^2_{\text{\tiny${{J/\psi}}$}}}}\frac{\Lambda_T^4}{4}\\
&(\Delta_3-2\Delta_4+\Delta_5)+\frac{1}{{m^2_{\text{\tiny${{ D^*}}$}}}{m^2_{\text{\tiny${{J/\psi}}$}}}}(\frac{\Lambda_T^6\Delta_6-2\Lambda_T^6 \Delta_7+2{\Lambda_T^4}}{8})+(\frac{\Lambda_T^2}{2}\Delta_1-\Lambda_T^2\Delta_2  +p^2)\frac{2p^2}{{m^2_{\text{\tiny$D_{(s)}^*$}}}}\\
&-\frac{p^4 \Delta_5}{{m^2_{\text{\tiny${{ D^*}}$}}}{m^2_{\text{\tiny${{J/\psi}}$}}}}(\frac{\Lambda_T^4 \Delta_3-2{\Lambda_T^4 \Delta_4}+2{\Lambda_T^2\Delta_1p^2}}{4})-\frac{2}{m^2_{\text{\tiny$J/\psi$}}}(\frac{\Lambda_T^4 \Delta_5}{4}+p^4 -\Lambda_T^2\Delta_2p^2)-\frac{1}{{m^2_{\text{\tiny${{ D^*}}$}}}{m^2_{\text{\tiny${{J/\psi}}$}}}}\\
&  (\frac{\Lambda_T^6 \Delta_6}{4}-\frac{\Lambda_T^4 \Delta_4p^2}{2}+\frac{\Lambda_T^4 \Delta_5 p^2}{2}-\frac{\Lambda_T^6 \Delta_7}{4}) +\frac{1}{{m^2_{\text{\tiny${{ D^*}}$}}}{m^2_{\text{\tiny${{J/\psi}}$}}}}
(\frac{\Lambda_T^6 \Delta_6}{8}  - \frac{\Lambda_T^4 \Delta_4p^2}{2}+\frac{\Lambda_T^2 \Delta_1p^4}{2})\Big),
\end{aligned}
\end{equation}
with $
\Delta=\frac{-m_{\text{\tiny$D^*$}}}{m_{\text{\tiny$\eta_c$}}+m_{\text{\tiny$D^*$}}}p^2+\alpha\beta m_{\text{\tiny$D^*$}}^2+(1-\alpha)\beta( p^2-  m_{\text{\tiny$\eta_c$}}^2)-\frac{1}{4}\frac{(2(1-\alpha)\beta-\frac{2m_{\text{\tiny$D^*$}}}{m_{\text{\tiny$\eta_c$}}+m_{\text{\tiny$D^*$}}})^2}{-1-\beta}p^2$, we take $\Lambda_T=1$ GeV in the work.
\begin{eqnarray}
\Delta^{\prime}&=&-\frac{m^2_{\text{\tiny$D^*$}}}{m_{\text{\tiny$J/\psi$}}+m_{\text{\tiny$D^*$}}}p^2+\alpha\beta m_{\text{\tiny$D^*$}}^2-(1-\alpha)\beta p^2+(1-\alpha)\beta  m_{\text{\tiny$J/\psi$}}^2+\frac{\Big(2(1-\alpha)\beta-\frac{-2m_{\text{\tiny$D^*$}}}{m_{\text{\tiny$J/\psi$}}+m_{\text{\tiny$D^*$}}}\Big)^2}{4(1+\beta)}p^2,\nonumber\\
\Delta_1&=&\frac{\left(2 \times(1-\alpha) \beta+\frac{2 m_{\text{\tiny$D^*$}}}{m_{\text{\tiny$J/\psi$}}+m_{\text{\tiny$D^*$}}}\right)^2 \Phi^2({\Delta^{\prime}})}{2(-\beta-1)^4 \Lambda_T^2}, \Delta_2=\frac{\left(-4 \times(1-\alpha) \beta-\frac{4 m_{\text{\tiny$D^*$}}}{m_{\text{\tiny$J/\psi$}}+m_{\text{\tiny$D^*$}}}\right) \Phi^2(\Delta^{\prime})}{2(-\beta-1)^3 \Lambda_T^2},\nonumber\\
\Delta_3&=&\Phi^2(\Delta^{\prime})\Big(\frac{3(2(1-\alpha)\beta+\frac{2m_{\text{\tiny$D^*$}}}{m_{\text{\tiny$J/\psi$}}+m_{\text{\tiny$D^*$}}})^2 }{(-\beta-1)^5\Lambda_T^2}+\frac{p^2(2(1-\alpha)\beta+\frac{2m_{\text{\tiny$D^*$}}}{m_{\text{\tiny$J/\psi$}}+m_{\text{\tiny$D^*$}}})^4  }{2(-\beta-1)^6\Lambda_T^4}\Big),\nonumber\\
\Delta_4&=& \Phi^2(\Delta^{\prime})\Big(\frac{3(4 \beta(\alpha-1) -\frac{4 m_{\text{\tiny$D^*$}}}{m_{\text{\tiny$J/\psi$}}+m_{\text{\tiny$D^*$}}})   }{2(\beta+1)^4 \Lambda_T^2}+\frac{p^2(4 (\alpha-1) \beta-\frac{4 m_{\text{\tiny$D^*$}}}{m_{\text{\tiny$J/\psi$}}+m_{\text{\tiny$D^*$}}})(2 \beta(1-\alpha)+\frac{2 m_{\text{\tiny$D^*$}}}{m_{\text{\tiny$J/\psi$}}+m_{\text{\tiny$D^*$}}})^2   }{2(-\beta-1)^5 \Lambda_T^4}\Big),\nonumber\\
\Delta_5&=&\Phi^2(\Delta^{\prime})\Big(\frac{(-4 \times(1-\alpha) \beta-\frac{4 m_{\text{\tiny$D^*$}}}{m_{\text{\tiny$J/\psi$}}+m_{\text{\tiny$D^*$}}})   }{(-\beta-1)^3 \Lambda_T^2}+\frac{p^2(-4 \times(1-\alpha) \beta-\frac{4 m_{\text{\tiny$D^*$}}}{m_{\text{\tiny$J/\psi$}}+m_{\text{\tiny$D^*$}}})  }{2(\beta+1)^4 \Lambda_T^4}\Big),\nonumber\\
\Delta_6&=&  \Phi^2(\Delta^{\prime})\Big(\frac{3}{(-\beta-1)^4 \Lambda_T^2}+\frac{2 p^2(-4 \times(1-\alpha) \beta-\frac{4 m_{V_1}}{m_{\text{\tiny$J/\psi$}}+m_{V_1}})^2 }{(-\beta-1)^5 \Lambda_T^4} + \frac{p^2(2 \times(1-\alpha) \beta+\frac{2m_{V_1}}{m_{\text{\tiny$J/\psi$}}+m_{V_1}})^2 }{(-\beta-1)^5 \Lambda_T^4}, \nonumber\\
& &+\frac{3 p^4(4 \times(\alpha-1) \beta-\frac{4 m_{\text{\tiny$D^*$}}}{m_{\text{\tiny$J/\psi$}}+m_{\text{\tiny$D^*$}}})^2(2 \times(1-\alpha) \beta+\frac{2 m_{\text{\tiny$D^*$}}}{m_{\text{\tiny$J/\psi$}}+m_{\text{\tiny$D^*$}}})^2 }{8(\beta+1)^6 \Lambda_T^6}\Big ),\nonumber\\
\Delta_7&=& \Phi^2(\Delta^{\prime})\Big( \frac{12p^2\alpha \beta-12p^2 \beta-12p^2 \frac{m_{\text{\tiny$D^*$}}}{m_{\text{\tiny$J/\psi$}}+m_{\text{\tiny$D^*$}}})}{(\beta+1)^4 \Lambda_T^4}+\frac{3p^4((\alpha-1) \beta-\frac{ m_{\text{\tiny$D^*$}}}{m_{\text{\tiny$J/\psi$}}+m_{\text{\tiny$D^*$}}})^3 }{2(-\beta-1)^5\Lambda_T^6}\Big).
\end{eqnarray}
The fully production and decay channels about four-quark molecular states directly are collected into Tab.\,\ref{tab:production} and Tab.\,\ref{tab:decay}. In addition, The complete relations between different production processes are listed,
\begin{eqnarray}
	&&\Gamma(B_c^+\to \rho^0  T_{J/\psi D^*}^{+})= \frac{1}{2}\Gamma(B_c^+\to \rho^+  T_{J/\psi D^*}^{0})= \frac{1}{2}\Gamma(B_c^+\to \overline{K}^{*0} T_{J/\psi D_s^*}^{+}),\nonumber\\
	&& \Gamma(B_c^+\to K^{*+}  T_{J/\psi D^*}^{0})= \Gamma(B_c^+\to K^{*0}  T_{J/\psi D^*}^{+})=\Gamma(B_c^+\to \phi T_{J/\psi D_s^*}^{+}),\nonumber\\
&&\Gamma(B_c^+\to \rho^0  T_{\eta_c D^*}^{+})= \frac{1}{2}\Gamma(B_c^+\to \rho^+  T_{\eta_c D^*}^{0})= \frac{1}{2}\Gamma(B_c^+\to \overline{K}^{*0} T_{\eta_c D_s^*}^{+}),\nonumber\\
	&& \Gamma(B_c^+\to K^{*+}  T_{\eta_c D^*}^{0})= \Gamma(B_c^+\to K^{*0}  T_{\eta_c D^*}^{+})=\Gamma(B_c^+\to \phi T_{\eta_c D_s^*}^{+}).
\end{eqnarray}
The relations between different decay channels are given as,
\begin{eqnarray}
 &&\Gamma(T_{J/\psi D^*}^{0}\to \eta_c  D^{*0})
= \Gamma(T_{J/\psi D^*}^{+}\to \eta_c  D^{*+})=\Gamma(T^{+}_{J/\psi D_s^*}\to \eta_c  D^{*+}_s),\,\,\,\nonumber\\
&&\Gamma(T_{J/\psi D^*}^{0}\to J/\psi  D^0)=\Gamma(T_{J/\psi D^*}^{+}\to J/\psi  D^+)=\Gamma(T_{J/\psi D_s^*}^{+}\to J/\psi  D^+_s),\,\,\,\nonumber\\
&&\Gamma(T^0_{\eta_c  D^{*0}}\to J/\psi  D^0)
= \Gamma(T^+_{\eta_c  D^{*0}}\to J/\psi  D^{+})=\Gamma(T_{\eta_c  D_s^*}^{+}\to \eta_c  D^{*+}_s),\,\,\,
\nonumber\\
&&\Gamma(T_{J/\psi D^*}^{0}\to    D^+  J/\psi  \pi^- )=\Gamma(T_{J/\psi D^*}^{0}\to    D^+_s  J/\psi  K^- )=\Gamma(T_{J/\psi D^*}^{+}\to    D^0  J/\psi  \pi^+ )\nonumber\\
&&=\Gamma(T_{J/\psi D^*}^{+}\to    D^+_s  J/\psi  \overline K^0)
=\Gamma(T_{J/\psi D_s^*}^{+}\to    D^0  J/\psi K^+ )=\Gamma(T_{J/\psi D_s^*}^{+}\to    D^+  J/\psi  K^0) \nonumber\\
&&=2\Gamma(T_{J/\psi D^*}^{0}\to  D^0   J/\psi   \pi^0  )=2\Gamma(T_{J/\psi D^*}^{+}\to    D^+  J/\psi  \pi^0)=6\Gamma(T_{J/\psi D^*}^{0}\to    D^0  J/\psi  \eta_q )\nonumber\\
&&=6\Gamma(T_{J/\psi D^*}^{+}\to    D^+  J/\psi  \eta_q )=\frac{3}{2}\Gamma(T_{J/\psi D_s^*}^{+}\to    D^+_s  J/\psi  \eta_q ),\,\,\,\Gamma(T_{J/\psi D^*}^{0}\to    D^+  \eta_c \pi^- )\nonumber\\
 &&=\Gamma(T_{J/\psi D^*}^{0}\to    D^+_s \eta_c  K^- )=\Gamma(T_{J/\psi D^*}^{+}\to    D^0 \eta_c  \pi^+ )=\Gamma(T_{J/\psi D^*}^{+}\to    D^+_s  \eta_c  \overline K^0)
\nonumber\\
 &&=\Gamma(T_{J/\psi D_s^*}^{+}\to    D^0  \eta_c K^+ )=\Gamma(T_{J/\psi D_s^*}^{+}\to    D^+  \eta_c  K^0) =2\Gamma(T_{J/\psi D^*}^{0}\to  D^0  \eta_c  \pi^0  )\nonumber\\
 &&=2\Gamma(T_{J/\psi D^*}^{+}\to    D^+  \eta_c \pi^0)=6\Gamma(T_{J/\psi D^*}^{0}\to    D^0  \eta_c  \eta_q )=6\Gamma(T_{J/\psi D^*}^{+}\to    D^+  \eta_c  \eta_q )\nonumber\\
 &&=\frac{3}{2}\Gamma(T_{J/\psi D_s^*}^{+}\to    D^+_s \eta_c  \eta_q ),\,\,\,\Gamma(T_{J/\psi D^*}^{0}\to    D^+  \eta_c \pi^- )=\Gamma(T_{J/\psi D^*}^{0}\to    D^+_s \eta_c  K^- )\nonumber\\
 &&=\Gamma(T_{J/\psi D^*}^{+}\to    D^0 \eta_c  \pi^+ )=\Gamma(T_{J/\psi D^*}^{+}\to    D^+_s  \eta_c  \overline K^0)
=\Gamma(T_{J/\psi D_s^*}^{+}\to    D^0  \eta_c K^+ )\nonumber\\
 &&=\Gamma(T_{J/\psi D_s^*}^{+}\to    D^+  \eta_c  K^0) =2\Gamma(T_{J/\psi D^*}^{0}\to  D^0  \eta_c  \pi^0  )=2\Gamma(T_{J/\psi D^*}^{+}\to    D^+  \eta_c \pi^0)\nonumber\\
 &&=6\Gamma(T_{J/\psi D^*}^{0}\to    D^0  \eta_c  \eta_q )=6\Gamma(T_{J/\psi D^*}^{+}\to    D^+  \eta_c  \eta_q )=\frac{3}{2}\Gamma(T_{J/\psi D_s^*}^{+}\to    D^+_s \eta_c  \eta_q ).
\end{eqnarray}
\begin{table}
	\caption{{The productions of the molecular states  $T_{cc\bar c \bar q}$  from $B_c$ meson. $T_{cc\bar c\bar q}$ can be molecule $T_{\eta_c D^*}$ and $T_{J/\psi D^*}$.}}\label{tab:production}\begin{tabular}{cccccc}\hline\hline
		channel & amplitude&channel & amplitude \\\hline

$B_c^+\to   \rho^0    T_{cc\bar c\bar d}^{+} $ & $ \frac{-a_1 V^*_{\text{cd}}}{\sqrt{2}}$
&$B_c^+\to   K^{*+}  T_{cc\bar c\bar u}^{0} $ & $ a_1 V^*_{\text{cs}}$\\
$B_c^+\to   \overline{K}^{*0}  T_{cc\bar c\bar d}^{+} $ & $ a_1 V^*_{\text{cs}}$&
$B_c^+\to   \overline{K}^{*0}  T_{cc\bar c\bar s}^{+} $ & $ a_1 V^*_{\text{cd}}$
\\$B_c^+\to   \rho^+    T_{cc\bar c\bar u}^{0} $ & $ a_1 V^*_{\text{cd}}$&$B_c^+\to   \phi  T_{cc\bar c\bar s}^{+} $ & $ a_1 V^*_{\text{cs}}$\\\hline
		\hline
	\end{tabular}
\end{table}
\begin{table}
\caption{Two-body and three-body decay processes  of $T_{J/\psi D^*}$ and $T_{\eta_c D^*}$. }\label{tab:decay}\begin{tabular}{ccccccc|c}\hline\hline
channel & amplitude & channel & amplitude &channel & amplitude \\\hline
$T_{J/\psi D^*}^{0}\to  J/\psi  D^0   $ & $ \frac{b_1}{3}$
&$T_{J/\psi D^*}^{+}\to   J/\psi D^+  $ & $ \frac{b_1}{3}$
&$T_{J/\psi D_s^*}^{+}\to  J/\psi  D^+_s   $ & $ \frac{b_1}{3}$\\
$T_{J/\psi D^*}^{0}\to   \eta_c D^{*0}   $ & $ \frac{b_2}{3}$
&$T_{J/\psi D^*}^{+}\to  \eta_c  D^{*+}   $ & $ \frac{b_2}{3}$
&$T_{J/\psi D_s^*}^{+}\to   \eta_c D^{*+}_s   $ & $ \frac{b_2}{3}$\\
$T_{\eta_c D^*}^{0}\to   J/\psi D^0   $ & $ \frac{b_1^{\prime}}{3}$
&$T_{\eta_c D^*}^{+}\to  J/\psi  D^+   $ & $ \frac{b_1^{\prime}}{3}$
&$T_{\eta_c D_s^*}^{+}\to  J/\psi  D^+_s  $ & $ \frac{b_1^{\prime}}{3}$\\\hline
$T_{J/\psi D^*}^{0}\to  D^0   J/\psi   \pi^0  $ & $ \frac{c_1}{\sqrt{2}}$&
$T_{J/\psi D^*}^{0}\to    D^0  J/\psi  \eta_q  $ & $ \frac{c_1}{\sqrt{6}}$&
$T_{J/\psi D^*}^{0}\to    D^+  J/\psi  \pi^-  $ & $ c_1$\\
$T_{J/\psi D^*}^{0}\to    D^+_s  J/\psi  K^-  $ & $ c_1$&
$T_{J/\psi D^*}^{+}\to    D^0  J/\psi  \pi^+  $ & $ c_1$&
$T_{J/\psi D^*}^{+}\to    D^+  J/\psi  \pi^0  $ & $ -\frac{c_1}{\sqrt{2}}$\\
$T_{J/\psi D^*}^{+}\to    D^+  J/\psi  \eta_q  $ & $ \frac{c_1}{\sqrt{6}}$&
$T_{J/\psi D^*}^{+}\to    D^+_s  J/\psi  \overline K^0  $ & $ c_1$&
$T_{J/\psi D_s^*}^{+}\to    D^0  J/\psi  K^+  $ & $ c_1$\\
$T_{J/\psi D_s^*}^{+}\to    D^+  J/\psi  K^0  $ & $ c_1$&
$T_{J/\psi D_s^*}^{+}\to    D^+_s  J/\psi  \eta_q  $ & $ -\sqrt{\frac{2}{3}} c_1$&
$T_{J/\psi D^*}^{0}\to    D^0  \eta_c  \pi^0  $ & $ \frac{c_2}{\sqrt{2}}$\\
$T_{J/\psi D^*}^{0}\to    D^0  \eta_c  \eta_q  $ & $ \frac{c_2}{\sqrt{6}}$&
$T_{J/\psi D^*}^{0}\to    D^+  \eta_c  \pi^-  $ & $ c_2$&
$T_{J/\psi D^*}^{0}\to    D^+_s  \eta_c  K^-  $ & $ c_2$\\
$T_{J/\psi D^*}^{+}\to    D^0  \eta_c  \pi^+  $ & $ c_2$&
$T_{J/\psi D^*}^{+}\to    D^+  \eta_c  \pi^0  $ & $ -\frac{c_2}{\sqrt{2}}$&
$T_{J/\psi D^*}^{+}\to    D^+  \eta_c  \eta_q  $ & $ \frac{c_2}{\sqrt{6}}$\\
$T_{J/\psi D^*}^{+}\to    D^+_s  \eta_c  \overline K^0  $ & $ c_2$&
$T_{J/\psi D_s^*}^{+}\to    D^0  \eta_c  K^+  $ & $ c_2$&
$T_{J/\psi D_s^*}^{+}\to    D^+  \eta_c  K^0  $ & $ c_2$\\
$T_{J/\psi D_s^*}^{+}\to    D^+_s  \eta_c  \eta_q  $ & $ -\sqrt{\frac{2}{3}} c_2$\\
$T_{\eta_c D_s^*}^{0}\to    D^0  \eta_c  \eta_q  $ & $ \frac{c_1^{\prime}}{\sqrt{6}}$&
$T_{\eta_c D^*}^{0}\to    D^+  \eta_c  \pi^-  $ & $ c_1^{\prime}$&
$T_{\eta_c D^*}^{0}\to    D^+_s  \eta_c  K^-  $ & $ c_1^{\prime}$\\
$T_{\eta_c D^*}^{+}\to    D^0  \eta_c  \pi^+  $ & $ c_1^{\prime}$&
$T_{\eta_c D^*}^{+}\to    D^+  \eta_c  \pi^0  $ & $ -\frac{c_1^{\prime}}{\sqrt{2}}$&
$T_{\eta_c D^*}^{+}\to    D^+  \eta_c  \eta_q  $ & $ \frac{c_1^{\prime}}{\sqrt{6}}$\\
$T_{\eta_c D^*}^{+}\to    D^+_s  \eta_c  \overline K^0  $ & $ c_1^{\prime}$&
$T_{\eta_c D_s^*}^{+}\to    D^0  \eta_c  K^+  $ & $ c_1^{\prime}$&
$T_{\eta_c D_s^*}^{+}\to    D^+  \eta_c  K^0  $ & $ c_1^{\prime}$\\
$T_{\eta_c D_s^*}^{0}\to    D^0  \eta_c  \pi^0  $ & $ \frac{c_1^{\prime}}{\sqrt{2}}$&
$T_{\eta_c D_s^*}^{+}\to    D^+_s  \eta_c  \eta_q  $ & $ -\sqrt{\frac{2}{3}} c_1^{\prime}$\\\hline
\hline
\end{tabular}
\end{table}

\end{document}